# Electric propulsion reliability: statistical analysis of on-orbit anomalies and comparative analysis of electric versus chemical propulsion failure rates

Joseph Homer Saleh*, Fan Geng, Michelle Ku, Mitchell L. R. Walker II

*School of Aerospace Engineering, Georgia Institute of Technology, Atlanta, GA 30332, USA*

**Abstract**: With a few hundred spacecraft launched to date with electric propulsion (EP), it is possible to conduct an epidemiological study of EP's on orbit reliability. The first objective of the present work was to undertake such a study and analyze EP's track record of on orbit anomalies and failures by different covariates. The second objective was to provide a comparative analysis of EP's failure rates with those of chemical propulsion. Satellite operators, manufacturers, and insurers will make reliability- and risk-informed decisions regarding the adoption and promotion of EP on board spacecraft. This work provides evidence-based support for such decisions. After a thorough data collection, 162 EP-equipped satellites launched between January 1997 and December 2015 were included in our dataset for analysis. Several statistical analyses were conducted, at the aggregate level and then with the data stratified by severity of the anomaly, by orbit type, and by EP technology. Mean Time To Anomaly (MTTA) and the distribution of the time to (minor/major) anomaly were investigated, as well as anomaly rates.

The important findings in this work include the following: (1) Post-2005, EP's reliability has outperformed that of chemical propulsion; (2) Hall thrusters have robustly outperformed chemical propulsion, and they maintain a small but shrinking reliability advantage over gridded ion engines. Other results were also provided, for example the differentials in MTTA of minor and major anomalies for gridded ion engines and Hall thrusters. It was shown that: (3) Hall thrusters exhibit minor anomalies very early on orbit, which might be indicative of infant anomalies, and thus would benefit from better ground testing and acceptance procedures; (4) Strong evidence exists that EP anomalies (onset and likelihood) and orbit type are dependent, a dependence likely mediated by either the space environment or differences in thrusters duty cycles; (5) Gridded ion thrusters exhibit both infant and wear-out failures, and thus would benefit from a reliability growth program that addresses both these types of problems.

* Corresponding author. E-mail address: jsaleh@gatech.edu (J.H. Saleh).

## 1. Introduction

The adoption of electric propulsion (EP) on board satellites has slowly but steadily increased over the last two decades. For example, a few hundred spacecraft have been launched to date with EP, and there are currently 128 active satellites in geosynchronous orbits (GEO) that use electrical propulsion (as of December 2015, excluding those with electro-thermal devices), up from 28 such satellites in 2000. Electric propulsion is also used on-board spacecraft in all other orbits including interplanetary orbits. EP is principally used for station-keeping and orbit repositioning, tasks with little acceleration demands, in conjunction with traditional chemical propulsion for orbit-raising. The situation however may slowly change in the near future as low thrust orbit changes from geo-transfer orbits (GTO) to GEO have already been demonstrated, and at least one major orbit-raising from a transfer orbit to GEO using only electric propulsion has already been completed (in 2011, with the Advanced Extremely High Frequency 1 satellite, following the failure of its main Liquid Apogee Engine). Furthermore, a recent announcement in 2012 by a major satellite manufacturer of the introduction of a fully electric satellite platform, and the acquisition of four of these platforms by a couple of satellite operators is an important new milestone for the EP technologies as it signals a changing attitude in the space community, traditionally and understandably risk averse, in its reliance on electric propulsion.

These recent developments are better appreciated when contrasted with the very slow, and at times hesitant history of development of EP. For instance, it is interesting to note that the inception of electric propulsion for space flight goes back a long way to Goddard in 1906:

> *"[Goddard] experimented with an electric gas discharge in 1906. As he observed the very high velocities which were imparted to the charged particles while the temperature of the tube remained fairly low, the thought occurred to him that electrostatically repelled particles might be the answer to the problem of obtaining high exhaust velocities at bearable chamber temperature [...]. The frequent recurrence of remarks concerning electrostatic propulsion in his notebooks from 1906 to 1912 reveals that ion [propulsion] had taken a firm foothold in [his] thinking*". Stuhlinger (1964)

The idea of electric propulsion was also proposed more than a century ago by Tsiolkovsky in 1911:

> "*It is possible that in time, we may use electricity to produce a large velocity for particles ejected from a rocket device [...]. It is known at present that cathode rays [...] are accompanied by a flux of electrons [...], the velocity of which are 6,000 to 20,000 times greater than that of the ordinary products of combustion*." (quoted in Choueri, 2004)

Little however was done with this idea of a "jet of charged particles" until Hermann Oberth (1894–1989) gave it a boost in 1929. In discussing the "electric spaceship", Oberth identified one of the most important advantages of EP, namely the mass-saving it provides. Details of this early history of EP can be found in Stuhlinger (1964, first chapter) and Choueri (2004). More than 50 years after its inception and following the dawn of the space age with the launch of Sputnik in 1957, EP was ready for its maiden flight. The first technology demonstrations and operational experiences with electric propulsion on orbit occurred in the 1960's both in the U.S. and the Soviet Union. In the U.S., NASA and the U.S. Air Force tested electric thrusters on board spacecraft between 1962 and 1971 (first with the Space Electric Rocket Test, SERT-1 suborbital mission, then with the Applications Technology Satellites, ATS-4 and ATS-5 for NASA, and with three SCOUT missions for the Air Force). In the Soviet Union, the probe Zond–2 was the first to use plasma thrusters in 1964.

After these early experiments, enthusiasm for electric propulsion in the U.S. seems to have dwindled, and there was a 20+ year hiatus before the Air Force resumed on orbit experimentation with EP thrusters. NASA continued to test EP on orbit but with a meager half a dozen flight experiments over the next two decades. The situation in the U.S.S.R. however was different, the enthusiasm for EP persisted and about 40 flights carried electric thrusters during the two-decade hiatus of the Air Force. As a result, EP matured earlier in the Soviet Union, and this may explain in part some of the reliability implications that will be seen later in this work. Japan also began experimenting with EP on orbit in the late 1970's and early 1980's, and was soon followed by China and Europe. Beyond its government support, commercial interest in EP developed slowly in the early 1980's, with a launch rate in the low single digit per year until the mid 1990's. A detailed review of the flight experience with EP can be found in Pollard and Janson (1996). These launch rates are better understood when contrasted with the launch rate of non-EP spacecraft: an average of over a 100 such spacecraft were launched per year[1] between the early 1960's and 1990's (Hiriart and Saleh, 2010), thus making EP during that time period occupy a very small niche market compared with chemical propulsion (CP).

It is against this background that the observations in the first paragraph in this Introduction have to be understood, namely that the late 1990's truly represent an inflection point in the adoption of EP, and the fact that the 128 satellites in GEO currently use EP is an important achievement for this 100+ year old idea.

It is worth reflecting, even if briefly, on the reasons for this sluggish development and belated market adoption of EP—technologies over a century in the making and with over 50 years of flight experience. Understanding the past can be informative about the extent and sustainability of the growth of EP adoption in the future:

1. First, with its minute thrust, EP was significantly behind the development priority of chemical propulsion before and early after the advent of the space age, and it carried little weight between the powerful advocates of liquid and/or versus solid propellant;

[1] The yearly launch rate fluctuated widely, and a sharp drop off for defense and intelligence spacecraft occurred in the late 1980's, from about 70 launches per year to about 30 per year in the early 1990's. A similar drop off occurred for science missions, albeit earlier starting in the 1970's. Commercial satellites exhibit a very different pattern, with a slow but steady growth until the mid 1990's (Hiriart and Saleh, 2010).

2. Second, given its level of thrust, EP was not suitable for operating in the atmosphere, and as such, it offered little appeal for weapon systems. Consequently, development funds for EP were minute compared with the interest in and support for chemical propulsion;
3. Third, one of the main advantages that EP provides over chemical propulsion, namely mass savings, had a low valuation in government procurement of spacecraft, until the late 1980's (given the geopolitical and military imperatives at the time). More generally, the advantages of EP did not seem appealing enough at the time to outweigh the drawbacks and technical uncertainties associated with it;
4. Fourth, the power available for spacecraft in terms of generation and storage (solar panels and batteries) was rather small (< 1 KW) and below a meaningful threshold for practical EP adoption;

All the above conspired synergistically to delay the development and adoption of EP and kept it stuck in the slow maturation lane. The situation began to change in the 1980's and 1990's for the following reasons:

5. In the 1980's and early 1990's, the mass savings afforded by EP were increasingly recognized as important in the commercial space market, not only for the cost savings they led to but also for the increased payload capacity which can replace the mass of propellant shaved. Given the large revenues communication satellites were reaping and the growing demand for their services, more payload capacity–as enabled by EP replacing chemical propulsion–would translate into more revenues. As a result, the value equation for EP began to change[2]. **New trade-offs would be enabled by the adoption of EP, including extending lifetime of the spacecraft on orbit and increasing its payload capacity, therefore modifying its value profile** (raising and extending the value delivery potential of the system);
6. While this value argument explains in part the inflection point in the adoption of EP in the 1990's, one reason for **the reluctance of its broader and enthusiastic adoption remained: the absence of a solid track record of on orbit performance and (demonstrated) field reliability**. Given the high cost of access to space and the quasi-unavailability of on orbit maintenance to compensate for subpar (hardware) reliability, **the risk aversion of the space community is understandable and it explains in part the slow uptake of EP even after the value argument won the day**.

With a few hundred spacecraft launched to date with electric propulsion, it is now possible to conduct an epidemiological study of EP's on-orbit track record of anomalies and failures. No such study has been conducted to date that included the global set of EP missions. The first objective of the present work is to undertake such a study and to analyze EP's track record of on orbit anomalies and failures (and identify different covariates). The second objective is to provide a comparative analysis of EP's (demonstrated) failure rate with that of chemical propulsion. Satellite operators, manufacturers, and insurers will make reliability- and risk-informed decisions regarding the adoption and insurance of EP on board spacecraft; this work provides evidence-based support for such decisions.

[2] A value analysis of EP would integrate the various benefits, costs, and drawbacks (including the much longer flight time to achieve final orbit and the corresponding revenues forfeited for example for a communications satellite), and benchmark the resulting net value against that of a system with chemical propulsion. Satellite operators will make value-informed decision regarding the adoption of EP, and it is important to understand under what conditions, and for what missions and markets, would EP tip the value balance in its favor. See Geng et al (2015), and Brathwaite and Saleh (2013) for examples of satellite value analysis in a commercial and government context.

The remainder of this work is organized as follows. In Section 2, we discuss our data and method. In Section 3, we conduct statistical analyses of EP anomalies and failures (by different covariates, including orbit, technology type, and severity). We also examine EP's time to anomaly, its Mean-Time-To-Anomaly by orbit type and technology (gridded ion engines versus Hall thrusters) as well as its distribution. Interesting patterns emerge from this analysis, and they lead to important insights (for EP testing purposes for example). In Section 4, we restrict our focus to satellites in GEO, and we analyze similar statistics for both EP and chemical propulsion. We then conduct a comparative analysis of failure rates of EP and chemical propulsion, and infer about the state-of-the-art of the reliability of one versus the other. In Section 5, we review the main findings and conclude this work.

## 2. Data and method

For the purpose of this study, as with our previous examination of spacecraft reliability (see for example Saleh and Castet, 2011; Kim et al., 2012), we relied to a large extent on the SpaceTrak database (Seradata 2016). This database is used by most of the world's launch providers, satellite insurers, operators, and manufacturers, and it provides extensive data on satellite on-orbit anomalies and failure as well as launch histories since 1957 (including launch manifest of several hundred payloads). SpaceTrak is considered the most authoritative database in the space industry with detailed information about launches, payloads, failures, and insurance losses for over 7,000 spacecraft. More details can be found in (Seradata, 2016).

This being said, we did not rely solely on the SpaceTrak database. Two additional steps were taken to ensure the validity and accuracy of our dataset before the analysis began: (1) we compared the SpaceTrak EP data with that of a major satellite insurer (for GEO satellites); and (2) we manually checked every spacecraft that remained in our dataset after the various data filters were applied (discussed next). The two datasets were to a large extent consistent over their common scope, and when they weren't, the manual check resolved the discrepancies and caught some minor issues: one spacecraft was miss-classified by orbit type, and seven EP spacecraft in GEO were missing. After this thorough data quality control, a total of 162 spacecraft launched between January 1997 and December 2015 remained in our data set. Basic descriptive summary statistics of this data set are provided at the end of this section and more details and analyses in sections 3 and 4.

### 2.1 Peculiarities and limitations of observational studies

Before discussing the different filters we applied for collecting the EP-specific data for this work, it is worth acknowledging upfront the peculiarities and limitations of statistical observational studies in general and this one in particular—a retrospective cohort study. Such studies are enabled by but also limited to the data collected and the level of resolution (or details) with which the data was recorded. For instance, anomalies in SpaceTrak are classified according to the severity of the event: a Class II anomaly for example is defined as a "major non-repairable failure that affects the operation of a satellite or its subsystems on a permanent basis." It is useful to have such a category for the anomaly data as it adds more resolution to the analysis beyond the traditional binary or dichotomous outcome (operational and failed, typical in many reliability studies). However, this class of "major" anomaly data does not have further level of resolution to quantify the extent of the permanent degradation suffered by the spacecraft. Hence this and similar studies are enabled by but also confined to the level of resolution with which the data was collected and recorded.

Another peculiarity of observational studies is that while valuable statistical associations and trends at the population level can be identified—and have important consequences for the design, redundancy allocation choices, and subsystem-level testing for example—very little physics of failure information can be harvested from such studies. To take a public health analogy for this situation, an observational study of air quality and health outcome in a community for example can identify and quantify the extent of an association between the exposure to varying air qualities and some disease, say the prevalence of asthma. The results can be very informative and useful in many ways, but the study cannot uncover the biological/etiological basis for the connection between air quality (e.g., aerosol) and asthma.

One last peculiarity worth pointing out is that field performance and reliability analysis is different from laboratory testing of technologies and subsystems. Both approaches are important and complementary, but different in spirit and objectives, and each has its own limitations. The latter is done in the controlled environment of a laboratory, with the subsystem usually not integrated into the overall system, and using a single or a small number of units to be tested. The objectives of lab testing are to assess the performance and qualify the technology for use in a particular environment, to identify its failure modes, and to assess the degradation path toward these failure modes. An important body of literature exist on testing and space-qualifying EP thrusters. For example, Delgado et al. (2014) discuss the qualification of the Hall effect thruster SPT-140 for use on Western spacecraft. The authors used two qualification models of the thruster, one to undergo qualification and acceptance testing and has accumulated several thousand hours of operation, and one to undergo plume characterization and other performance testing. The performance evaluation of the second model is discussed in Garner et al. (2015) in which the authors conclude that the SPT-140 is a "viable candidate for NASA missions requiring power throttling down to low thruster input power". Dankanich et al. (2009) develop a qualification standard for electric thrusters, which have very different lifetime qualification issues than chemical thrusters. The authors discuss the distinctive challenges in qualifying EP thrusters, namely "an operational lifetime of tens of thousands of hours, operation over a broad range of input powers, and complex wear-out failure modes". **Long lifetime requirements**, the authors point out, **make extensive testing infeasible, and** they conclude that **reliability or failure distribution cannot be assessed based on testing alone** (especially with a small number of units tested). The authors recognize that "**traditional approach of performing a single life test for the required [design] lifetime plus some margin–typically 50%–provides insufficient information to characterize the failure risk for a given space mission**". To mitigate this fact, accelerated testing[3] is increasingly used to cover the many years of required EP lifetime and uncover failure modes in various thruster components (Dankanich et al., 2009). Goebel et al., (2009) discuss testing and evaluation of another EP technology, namely the 25-cm XIPS© thruster. The authors point out that the thruster had already "completed 16,250 hours of life test with 14,134 on-off cycles", but that deep space missions would require significantly longer operating times. As a result, they undertook to conduct further testing and forensic analysis of thruster and the mechanism responsible for the erosion and wear-out of its components (in particular the discharge and neutralizer cathodes, and the high voltage grids). So how is field performance and reliability analysis different from laboratory testing of technologies and subsystems?

Analysis of field performance, in our case on-orbit EP reliability, including field anomalies and failures, is different from laboratory testing in several ways:

1. The focus is typically on the collective behavior of many units, not just an individual or a small number of test models;
2. The ground testing environment is fundamentally different from operation in the space environment, and as a result, different failure modes might emerge and/or manifest themselves differently (e.g., sooner);
3. The subsystem is integrated and operates within the overall system, and consequently different issues might emerge from the interactions between the system and the subsystem (as well as the subsystem and its environment), which were not seen in ground testing and qualification of the technology[4].

The importance of field performance analysis is self-evident as it helps to verify, validate, and identify deficiencies and blind spots in laboratory testing. It also constitutes an important feedback loop for learning and improving design, testing, and operations.

The present work fits in this tradition of field data analysis. Four previous articles in this tradition are worth mentioning. The first by Casaregola (2015) in which the author discusses the flight experience of Eutelsat with electric propulsion. The article is focused on the experience of this particular satellite operator (Eutelsat) with its two GEO satellites equipped with electric propulsion. Although this is a very small sample (two case studies), the work is important as it provides detailed discussion of field performance of EP on board these two satellites, SeaSat 1 and Ka-Sat, and it concludes that EP has reached a level of maturity and reliability suitable for commercial applications. This is a useful data point, but it needs more robust examination before the extrapolation from two case studies can be generalized and found to be statistically significant. For example, the article notes that "no [on-orbit] critical [EP] anomalies on board 17 other satellites] were reported to the author's knowledge"; it will be shown shortly this statement is ill-informed as several major anomalies on-orbit have occurred with EP. The second article by Corey and Pidgeon (2009) discusses the experience of a satellite manufacturer, Space Systems/Loral, with electrical propulsion. The authors summarize the field performance (and hardware description) of SS/L 5 EP-equipped spacecraft (5 when the article was published, 9 others were under construction at the time of publication). The work focuses on the SPT-100 Hall thruster. Also worth mentioning is the article by Estublier et al., (2007). The authors discuss the on-orbit performance of the Hall thruster on-board the all-electric SMART–1 spacecraft of the European Space Agency. EP was used as the primary propulsion system and achieved a number of "firsts", including the first use of EP to escape Earth gravity (from a Geo Transfer Orbit, GTO) and the first use of EP to achieve capture and descent around a celestial body (the Moon). These articles (Casaregola, 2015; Corey and Pidgeon, 2009; Estublier et al., 2007), along with the excellent publication by Wade et al., (2015) in which the authors examine the insurance implications of all-electric satellites, are important contributions to the literature on field performance of EP. Our work is in their spirit; it differs and extends them in ways discussed next.

---

[3] "The only accelerated test method currently used in thruster qualification is to increase the duty cycle by shortening the off time in cycled tests. This is considered representative of actual operation as long as the proper thermal conditions at start up are reproduced." (Dankanich et al., 2009)

[4] Some testing includes the integrated system. Despite this, as noted in Estublier et al. (2007) when discussing the all-electric SMART–1 spacecraft, "**measurements in vacuum chambers were not very representative and the real spacecraft configuration could not be reproduced; flight experience was essential**." We recognize that acceptance testing in a commercial environment is different from the typical integrated testing in a NASA and ESA environment.

### 2.2 Data filters

We applied several filters to our data collection before settling on a dataset to analyze. First, we restricted the time window to spacecraft launched with EP after January 1, 1997. Our objective in doing so was to limit the confounding effect of older EP technologies, and to examine the reliability and failure behavior of more recent ones. Incidentally, XIPS© and SPT saw their first commercial use in the late 1990's. Second, we eliminated electro-thermal devices from our dataset (arcjets and resistojets). EP technologies are typically classified according to the way the propellant or plasma is accelerated: electro-thermal, electrostatic, or electromagnetic (Martinez-Sanchez and Pollard, 1998; Toki et al., 2002; Jahn and Choueiri, 2002):

- *"Electro-thermal propulsion*, wherein the propellant is heated by some electrical process, then expanded through a suitable nozzle;
- *Electrostatic propulsion*, wherein the propellant is accelerated by direct application of electrostatic forces to ionized particles;
- *Electromagnetic propulsion*, wherein the propellant is accelerated under the combined action of electric and magnetic fields." (Jahn and Choueiri, 2002)

We removed the electro-thermal devices from our analysis since they would render the dataset technologically heterogeneous, and the high prevalence of such devices on-orbit would mask or confound the failure behavior of the two other EP technologies. The on-orbit performance of electro-thermal devices has been examined to some extent in Hoskins et al., 2013. Third, SpaceTrak identifies four classes of failure events, listed next in increasing order of severity:

- Class IV: minor/temporary/repairable failure that does not have a significant permanent impact on the operation of the satellite or its subsystems;
- Class III: major non-repairable failure that causes the loss of redundancy to the operation of a satellite or its subsystems on a permanent basis;
- Class II: major non-repairable failure that affects the operation of a satellite or its subsystems on a permanent basis. This effectively means major/significant losses, but not total immediate loss of the satellite;
- Class I: subsystem failure causing satellite retirements. This effectively means immediate and total loss of the satellite.

We collected all EP related anomalies within the time window of interest, and for convenience, we limited our analysis to three categories of failure events, *minor* (Class IV and Class III), *major* (Class II), and *fatal* (Class I). The rationale for conflating Class IV and Class III is twofold: on the one hand, the information regarding the availability and extent of EP redundancy on-board satellites is not available, and on the other hand, both classes of events have limited or no permanent impact on the operation of the satellite.

After the filters were applied, we were left with a dataset of 162 EP-equipped spacecraft launched between January 1, 1997 and December 31, 2015 and a corresponding flight experience of 429,939 days or about 10 million flight-hours on orbit. Additional details of this dataset are provided in Table 1, and further descriptive statistics are provided in the next sections.

**Table 1**. Dataset of EP-equipped spacecraft launched between 1997 and 2015

| | | | | |
|---|---|---|---|---|
| Total number of EP-equipped spacecraft, N | **162** | | | |
| Total flight experience | **429.939 days**<br>(10,318,536 hours) | | | |
| Orbit type | *LEO* | *MEO* | *GEO* | *Interplanetary / Trans-lunar* |
| Total number of EP-equipped spacecraft in our sample | 12 | 2 | 142 | 6 |
| Number of EP anomalies, n | **39** | | | |

In the next section, we analyze this dataset and examine various statistics related to these anomalies by stratifying along different covariates including orbit, technology type, and severity of the failure event. We then restrict our analysis to a single orbit, GEO, and we examine the anomaly rates of EP over two time periods, 1997–2004 and 2005–2015, for reasons that will be discussed shortly. Finally, one important objective of this study is to provide a comparative analysis of EP's failure rate with that of chemical propulsion in GEO. To this end, we collected data for all satellites equipped with chemical propulsion in GEO over the same time period as well as their anomalies (basic statistics in Table 2). We provide in Section 4 a comparative analysis of failure rates of EP versus chemical propulsion in GEO, pre- and post-2005.

**Table 2**. Dataset for chemical propulsion (CP) spacecraft in GEO launched between 1997 and 2015

| | |
|---|---|
| Total number of CP spacecraft, N | **482** |
| Total flight experience | **1,442,192 days**<br>(34,612,608 hours) |
| Number of CP anomalies, n | **37** |

It is useful to clarify upfront that the total flight time in Tables 1 and 2 refers to the collective time accrued when a spacecraft is launched until it is retired or completely fails on orbit, or until it reaches the end of our observation window of December 31, 2015. Similarly, the calculations of Mean Time to Anomaly (MTTA) in Section 3 and the anomaly rates in Section 4 are related to this definition of flight time, and not to the duty cycle of the propulsion subsystem.

## 3. EP anomalies: on-orbit data analysis and results

In this section, we analyze the prevalence of EP anomalies by different covariates, including orbit and engine type as well as the severity of the anomaly event. We also examine EP's Mean Time To Anomaly (MTTA) and some associated statistics, including the distribution of the time to (minor/major) anomaly.

### 3.1 Prevalence of failure events by severity, and time to EP anomaly

The breakdown of EP anomalies by event severity is shown in Figure 1. The figure shows a prevalence of minor EP anomalies (64%), but also a substantial proportion of major non-repairable EP anomalies affecting the spacecraft on a permanent basis (36%).

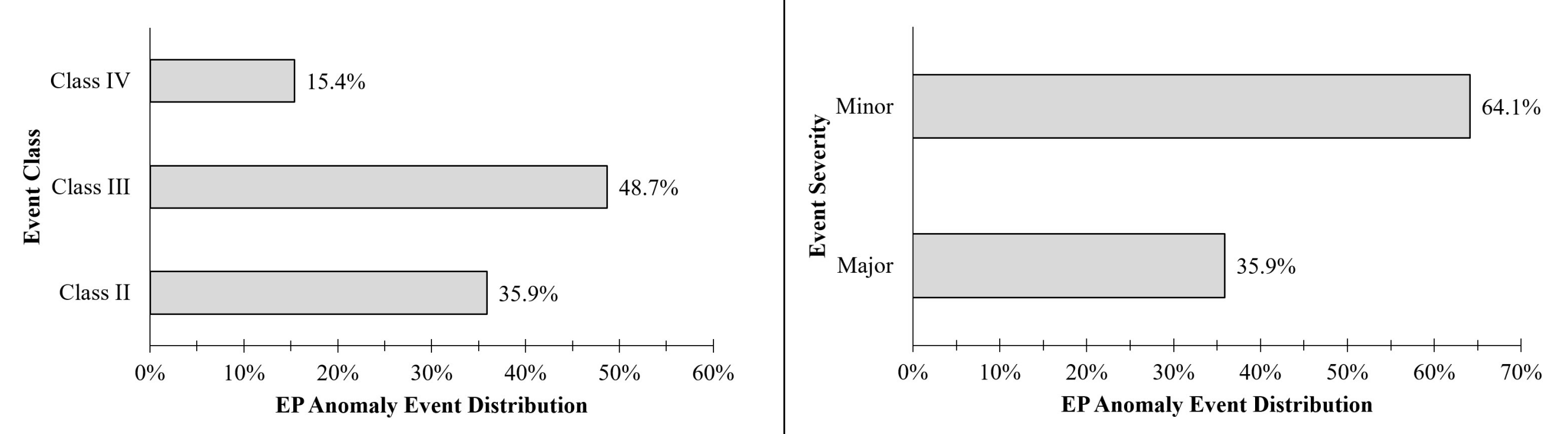

**Figure 1**: EP anomalies (all orbit types), classified by failure event Class and severity (n = 39). No Class I anomaly (fatal) occurred within the time period of interest (January 1, 1997 and December 31, 2015)

The Mean Time To (EP) Anomaly is provided in Figure 2, along with the 90% confidence intervals. The results show a significantly shorter MTTA for minor anomalies than for major anomalies, 758 days versus 1378 days. The results are even more striking for the median time to anomaly, a statistic that, unlike the mean, is not affected by outliers: about 304 days for the minor anomaly versus 1010 days for the major anomaly. The 90% confidence intervals are calculated as shown in (Eq. 1) using the *t* distribution with respectively 24 and 13 degrees of freedom for the minor ($n_{minor} = 25$) and major anomalies ($n_{major} = 14$). Given the small sample available, we choose the 90% confidence interval (CI) and the 10% significance level ($\alpha$) instead of the traditional 95% CI and $\alpha = 5\%$.

$$MTTA \pm t_{1-\frac{\alpha}{2}} \cdot \frac{stdev}{\sqrt{n_i}} \qquad (Eq. 1)$$

with

*stdev*: standard deviation of the sample

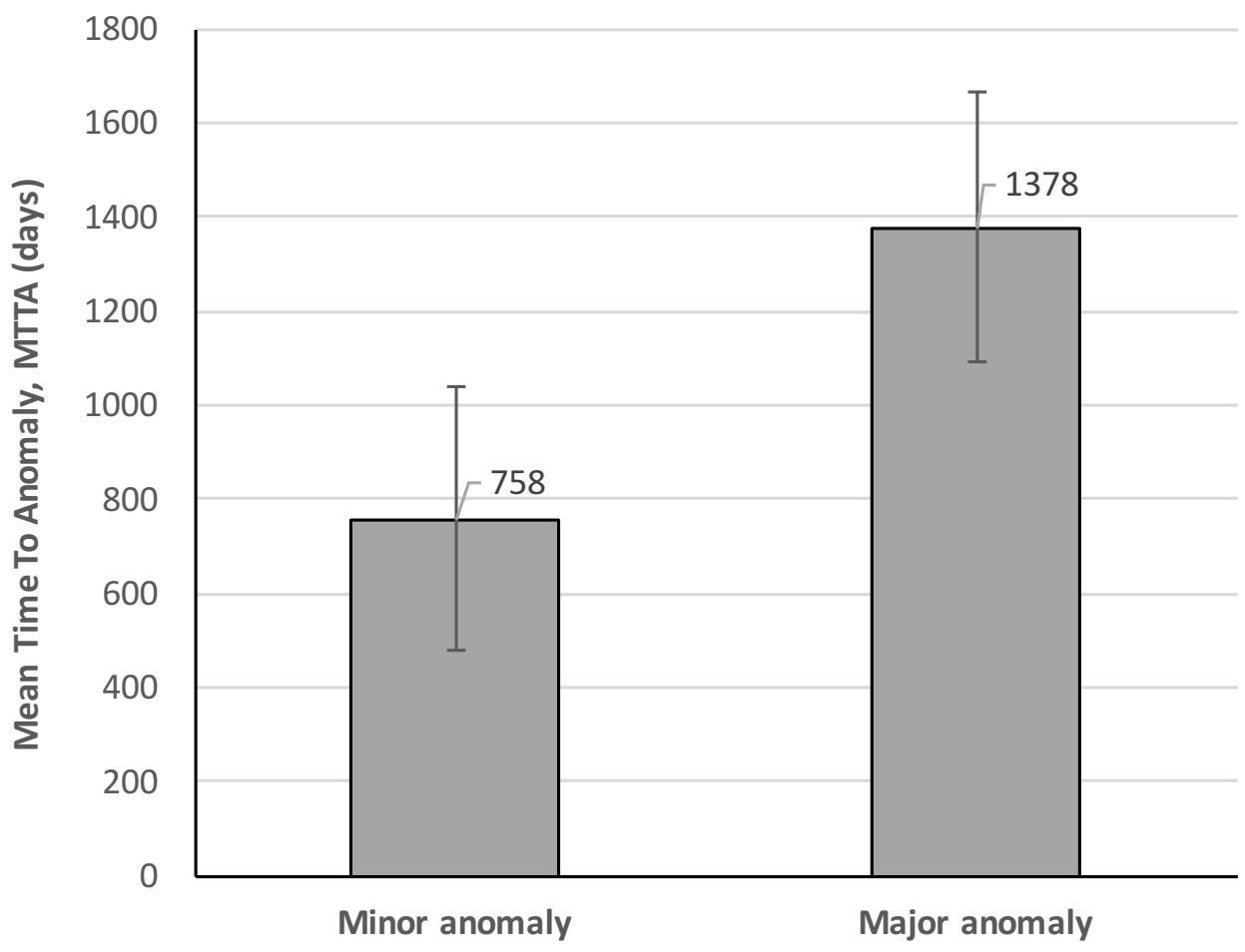

**Figure 2**. Mean Time To (EP) Anomaly by severity of all anomaly events, along with the 90% confidence intervals

The results for the MTTA show a statistically significant difference between the mean time to minor anomaly and mean time to major anomaly (p-value = 0.053, calculated using the two-sample mean comparison t-test). We further examined the distributions of the (random variable) time-to-anomaly for the major and minor categories by trying different probability plots, and found that these variables are roughly Weibull distributed, as shown in Eq. 2 and Figure 3.

$$\begin{cases} P\left(T_{anomaly} > t\right) = e^{-\left(\frac{t}{\theta}\right)^{\beta}} \\ \quad \beta: shape\ parameter \\ \quad \theta: scale\ parameter \end{cases} \qquad (Eq. 2)$$

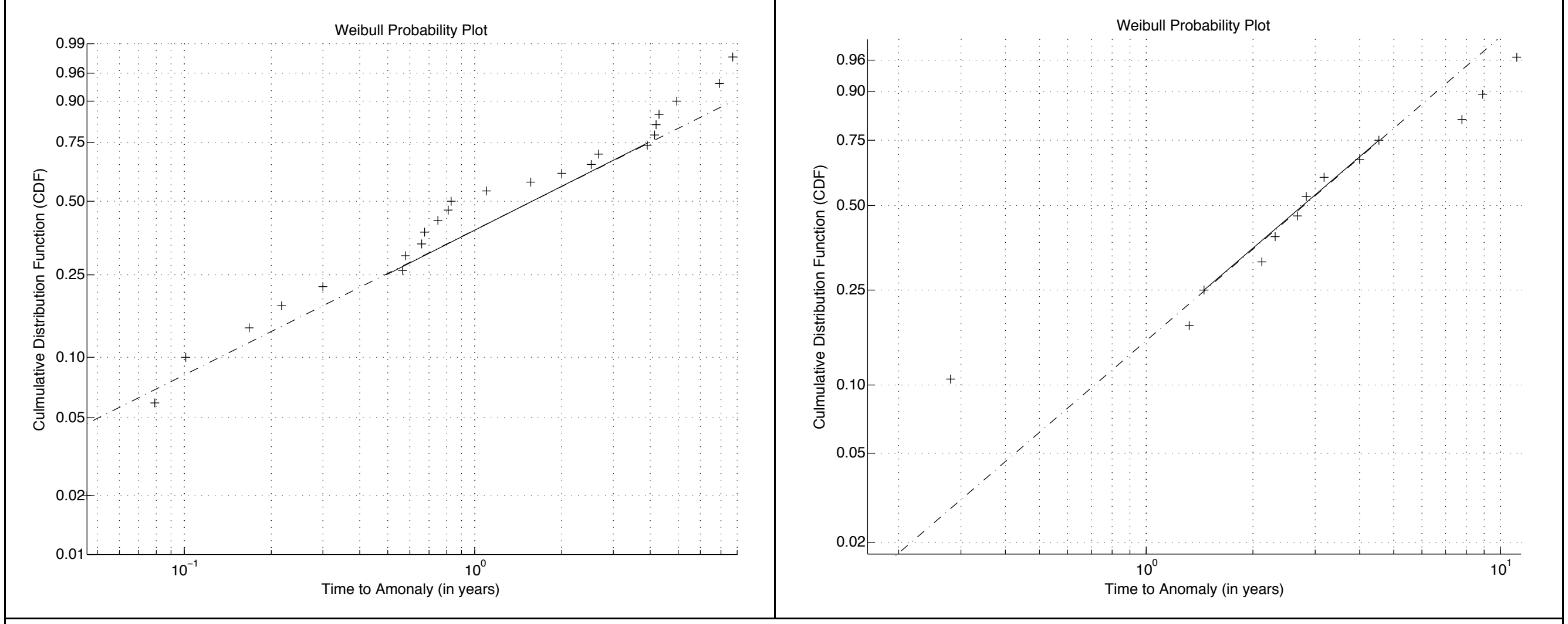

**Figure 3**: EP minor anomaly Weibull probability plot (left panel), and major anomaly Weibull probability plot (right)[5]

The scale and shape parameters of these distributions are provided in Table 3. The fact that the shape parameter (*β*) for the minor anomalies is less than 1 corresponds to a decreasing failure rate with time, and is indicative of the equivalent of infant mortality in reliability analysis, an important property of the Weibull distribution. This finding will be referred to as **EP's infant-minor-anomalies**. Such anomalies require better ground testing, improved quality screening, and dedicated burn-in procedures to address and eliminate. In contrast, the shape parameter for major anomalies is larger than 1, which corresponds to an increasing failure rate with time, and is indicative of wear-out failures (due to aging, fatigue, and/or erosion for example). This finding will be referred to as **EP's wear-out major anomalies**. Such anomalies are typically addressed by redundancy, de-rating, and ultimately better design. The calculated MTTA provided in Table 3 based on these Weibull probability plots are consistent with the empirical results in Figure 2 (subject to round-off errors and quality of fit).

**Table 3**. Scale and shape parameters for the Weibull distribution of the time-to- (minor, major) EP anomaly

| **Minor anomaly** | | **Comments** |
|---|---|---|
| Shape parameter, β | 0.86 | Indicative of infant mortality, or EP infant-minor-anomalies. Requires better ground testing, improved quality screening, and dedicated burn-in procedures to eliminate |
| Scale parameter, θ | 699 days | |

[5] Larger and clearer versions of these plots are provided in the appendix.

| | | |
|---|---|---|
| Calculated $MTTA_{minor}$ (Γ: gamma distribution) | $\theta \cdot \Gamma\left(1+\frac{1}{\beta}\right) = 755 \quad days$ | Compare with empirical $MTTA_{minor}$ of 758 days |
| **Major anomaly** | | |
| Shape parameter, β | 1.14 | Indicative of wear-out failures, or EP wear-out major anomalies. Typically addressed by redundancy, de-rating, and ultimately better design |
| Scale parameter, θ | 1442 days | |
| Calculated $MTTA_{major}$ | $\theta \cdot \Gamma\left(1+\frac{1}{\beta}\right) = 1376 \quad days$ | Compare with empirical $MTTA_{major}$ of 1378 days |

### 3.2 EP anomalies by orbit type: prevalence, test of independence, and time to anomaly

Figure 4 shows the prevalence of EP anomalies by orbit type and severity. These are descriptive statistics of the EP anomaly dataset, and they have to be cautiously interpreted since they are not normalized by the number of EP-equipped spacecraft in each orbit, that is, by all EP-equipped spacecraft, including those that have not experienced any EP anomaly. The salient results in Figure 4 are as follows: (1) the absence of EP anomalies in LEO/MEO; (2) the majority of EP anomalies are in GEO (66.6% of all EP anomalies occurred in GEO), and they are roughly equally split between minor and major; and (3) most of the EP anomalies in trans-lunar/interplanetary orbits are minor (84.6% of all EP anomalies in this type of orbit are minor).

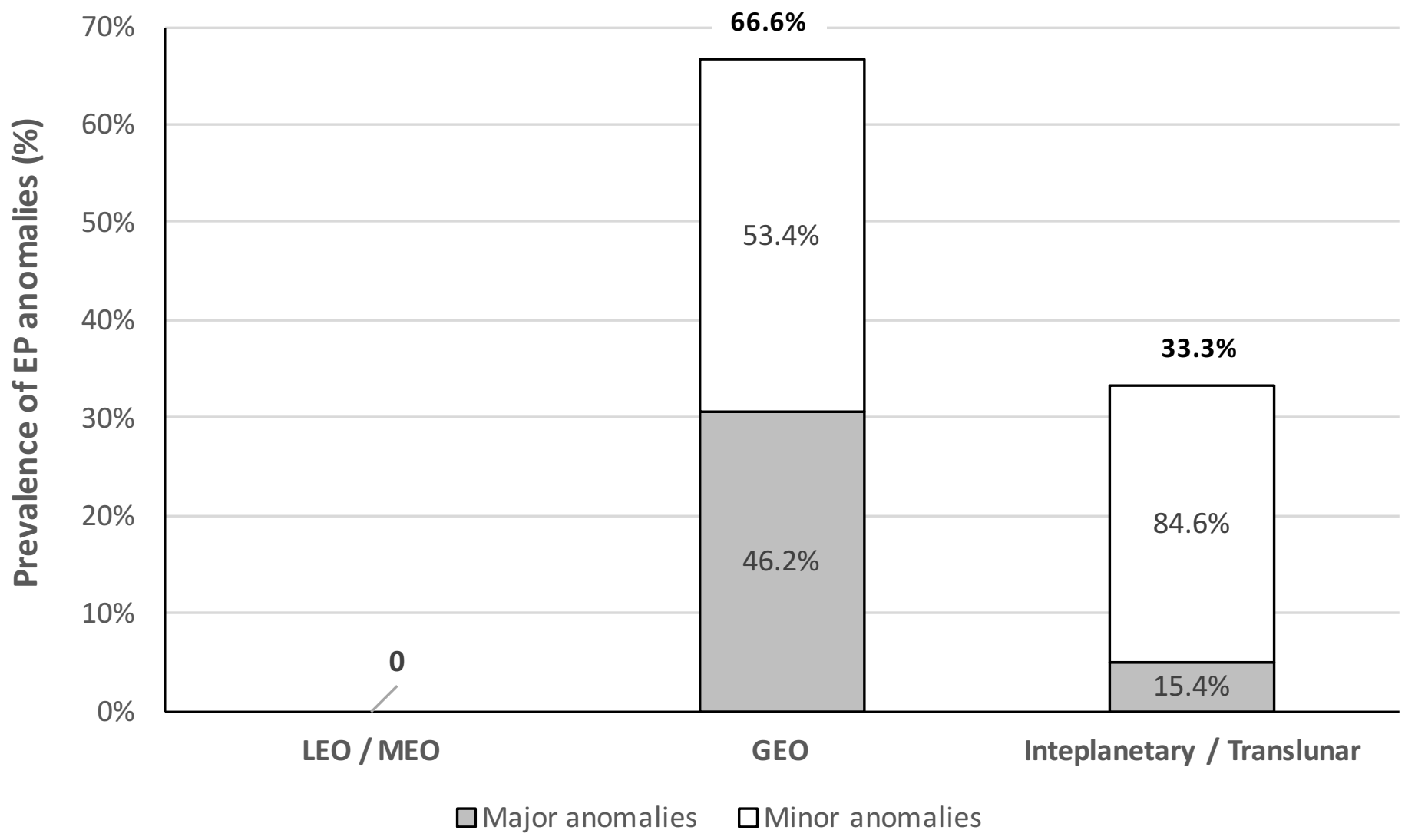

**Figure 4**. Prevalence of EP anomalies by orbit type and by severity (n = 39).

When normalized by the number of EP-equipped spacecraft per orbit (shown in the left-panel in Figure 5), the results show a significantly higher prevalence of anomalies per spacecraft in trans-lunar/interplanetary orbits than in GEO—about an order of magnitude difference (right-panel in Figure 5). These results however should not be interpreted as rates since *rate* in reliability and epidemiological studies intrinsically accounts for time in the denominator (which is not the case in these calculations). Anomaly rates will be carefully examined in Section 4.

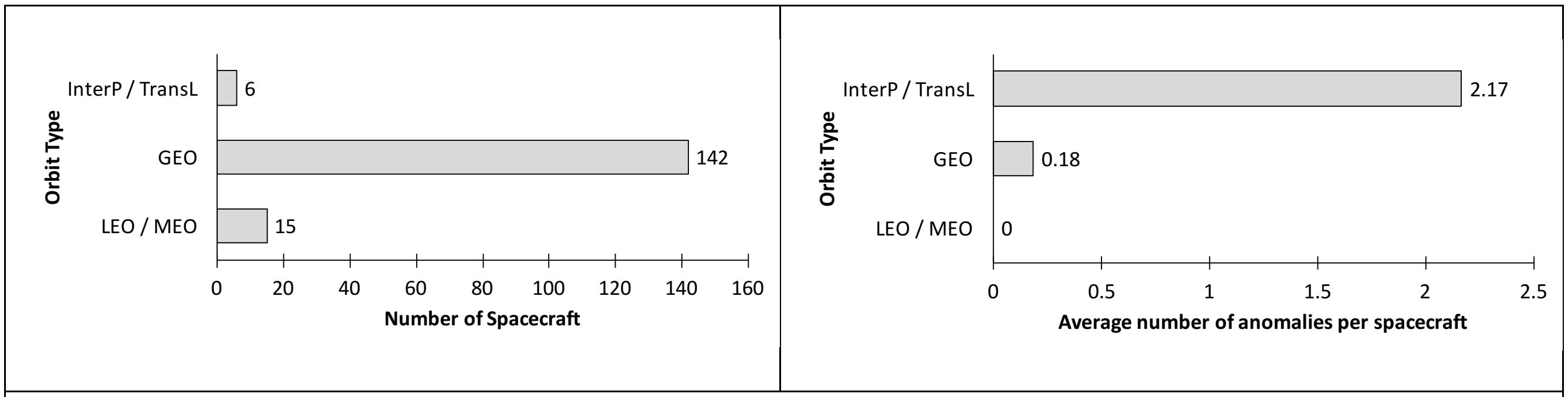

**Figure 5**: Total number of EP-equipped spacecraft by orbit (left panel), and average number of anomalies per spacecraft per obit (right)

The results in Figure 5 (right panel) are subject to sampling variability, and they prompted us to conduct a more formal chi-squared test of independence to assess whether the likelihood of a spacecraft to experience EP anomalies is independent of orbit type or not:

$H_0$: Likelihood of a spacecraft to experience EP anomalies is **independent** of orbit type
$H_a$: Likelihood of a spacecraft to experience EP anomalies is **dependent** on orbit type

The contingency table with *observed* and *expected frequencies* of EP anomalies are shown in Table 4. For a given cell *i,j* in the table, the observed ($O_{ij}$) and Expected ($E_{ij}$) counts of spacecraft with or without anomalies are provided. The chi-squared statistic is calculated as follows[6]:

$$\chi^2 = \sum_i \sum_j \frac{\left(O_{i,j} - E_{i,j}\right)^2}{E_{i,j}} = 30.4 \qquad (Eq.3)$$

With two degrees of freedom, the right-tail of the corresponding chi-squared distribution covers a p-value < 0.000001, thus providing **very strong evidence against the null hypothesis *$H_0$*. We therefore reject *$H_0$* and conclude that the likelihood of a spacecraft to experience EP anomalies and orbit type are dependent**.

[6] Minor point of statistical detail: The p value computed with the chi-squared test ($< 10^{-6}$) offers a comfortable margin for the conclusion. However, because two cells in Table 4 have expected counts less than 5, one of the assumptions of the chi-squared test is not satisfied, and as a result the p-value so computed may be biased. To overcome this situation, the "Fisher exact test" is the recommended alternative. No test statistic is calculated with the Fisher test, but a p-value < 0.0001 is directly provided. The conclusion is not affected by this more precise but slightly larger p-value.

While this test result only proves association not causation, it is fair to advance a causal hypothesis for this dependence, namely that the space environment mediates between orbit type and EP anomalies, and is likely one risk factor for the occurrence EP anomalies. This hypothesis deserves further investigation and constitutes a fruitful venue for future work. Another competing hypothesis is equally likely, namely that this association is mediated by excessive early usage or significantly higher duty cycle of EP thrusters onboard interplanetary/Trans-lunar orbits than on-board Earth-orbiting satellites[7]. Said differently, **two possible confounders for the results in Figure 5 and Table 4 are the differences in space environment and differences in duty cycle** (frequency and duration of EP usage, on/off cycles). Unfortunately, it is practically impossible to obtain the frequency and duration of the entirety of EP duty cycles from all the satellite operators involved in 162 spacecraft in our dataset. We leave this issue as a useful challenge for the interested researchers, to disentangle the effects of space environment and duty cycle on the likelihood of EP anomalies.

**Table 4**. Contingency table of spacecraft with and without EP anomalies by orbit type (parenthesis represent expected counts)

| | **LEO / MEO** | **GEO** | **Interplanetary / Trans-lunar** | Total |
|---|---|---|---|---|
| **Number of spacecraft with EP anomalies** | 0 *(1.73)* | 15 *(17.53)* | 5 *(0.74)* | 20 |
| **… without EP anomalies** | 14 *(12.27)* | 127 *(124.46)* | 1 *(5.26)* | 142 |
| Total | 14 | 142 | 6 | 162 |

The Mean Time To (minor/major) Anomaly in GEO are provided in Figure 6 along with the 90% confidence intervals. The same patterns in Figure 2 are again seen here with the data stratified by orbit type. Figure 6 also shows the The Mean Time To minor Anomaly in Interplanetary / Trans-lunar orbits. No statistically significant difference is found between the $MTTA_{minor}$ in GEO and Interplanetary / Trans-lunar orbits (this can be seen in the highly overlapping confidence intervals for the corresponding $MTTA_{minor}$). One limitation at this point should be acknowledged: as we restrict the data and control for more covariates (severity and orbit type), the sample size further shrinks (e.g., 12 major anomalies in GEO), the confidence intervals become larger, and it becomes increasingly more difficult to find statistically significant differences and association. For example, the major anomalies in Interplanetary / Trans-lunar orbits are not shown in Figure 6 because only two such events occurred (MTTA = 319 days) and the corresponding confidence interval is significantly large it is rather uninformative.

[7] We are grateful for an anonymous reviewer who pointed this out and suggested that "that the operating duty cycle of GEO (and most LEO/MEO) EP-equipped spacecraft (typically 0.1 or so) is very different than that of planetary spacecraft (typically >0.90 or so)".

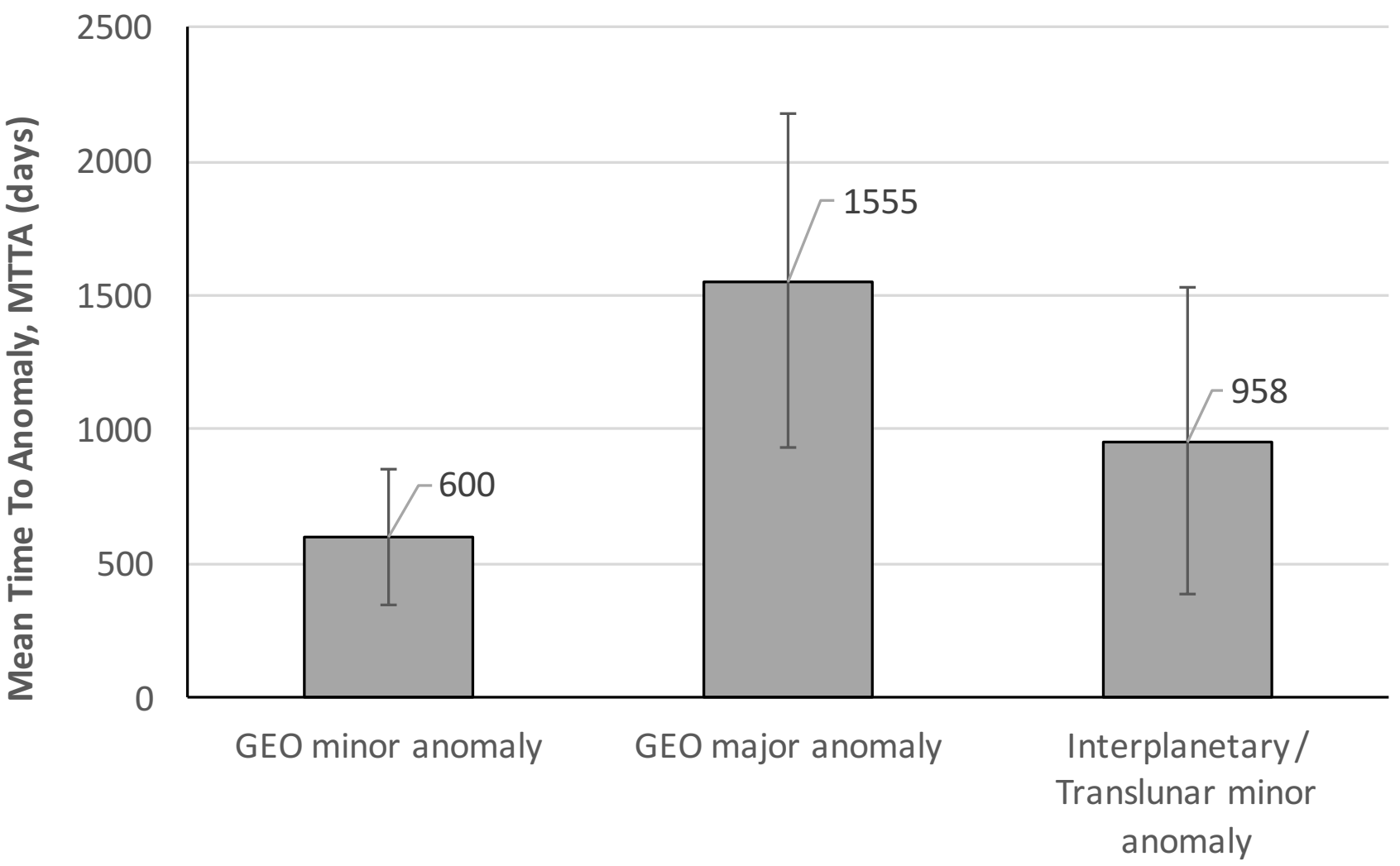

**Figure 6**. Mean Time To Anomaly in GEO and Interplanetary / Trans-lunar orbits, along with the 90% confidence intervals

Beyond the MTTA results, Table 5 provides the time for the onset of all EP anomalies as well as the first quartile EP anomalies in GEO and Interplanetary / Trans-lunar orbits. The results show that EP anomalies begin much sooner in Interplanetary / Trans-lunar orbits than in GEO (about 3.5 times sooner), and that the first quartile of all EP anomalies occur before 79 days in Interplanetary / Trans-lunar orbits, compared with the 297 days in GEO (about 3.7 times sooner). This results provides another perspective on the previous test of independence of EP anomalies and orbit types, namely that not only is the likelihood of EP anomalies and orbit type dependent, but also that the onset and early clustering of EP anomalies is also dependent on orbit type. Said more casually, EP anomalies are more likely and they begin sooner in Interplanetary / Trans-lunar orbits than in GEO. It is worth pointing out that these two characteristics, likelihood and onset with early clustering of EP anomalies, are not the same thing, and they might both be mediated by the differences in space environment between the two types of orbit or by early excessive usage on orbit (differences in duty cycles), as noted in the two previous causal hypotheses.

**Table 5**. Time for the onset of EP anomalies and first quartile, by orbit type

| | **GEO** | **Interplanetary / Trans-lunar** |
|---|---|---|
| **Onset of EP anomalies (days)** | 60 | 17 |
| **First quartile of EP anomalies (days)** | 297 | 79 |

### 3.3 EP anomalies by technology type: gridded ion engines versus Hall thrusters

Figure 7 shows the prevalence of EP anomalies by severity and by engine type, namely gridded ion engines and Hall thrusters. Details about these technologies can be found in Martinez-Sanchez and Pollard, 1998, and Jahn and Choueiri, 2002. These results as those in Figure 4 are descriptive statistics of the EP anomaly dataset, and they have to be cautiously interpreted since they are not normalized by the number of EP-equipped spacecraft with each type of engine, including those that have not experienced any EP related anomaly. The salient feature in Figure 7 is that most of the EP anomalies on orbit (~90%) have occurred in spacecraft equipped with gridded ion engines. Less salient but interesting to note is the fact that both types of engines experienced a majority of minor anomalies, 62.9% and 75% for the ion engines and Hall thrusters respectively. Given the small sample size, we can loosely say that ion engines and Hall thrusters have exhibited roughly the same ratio of minor to major anomalies[8].

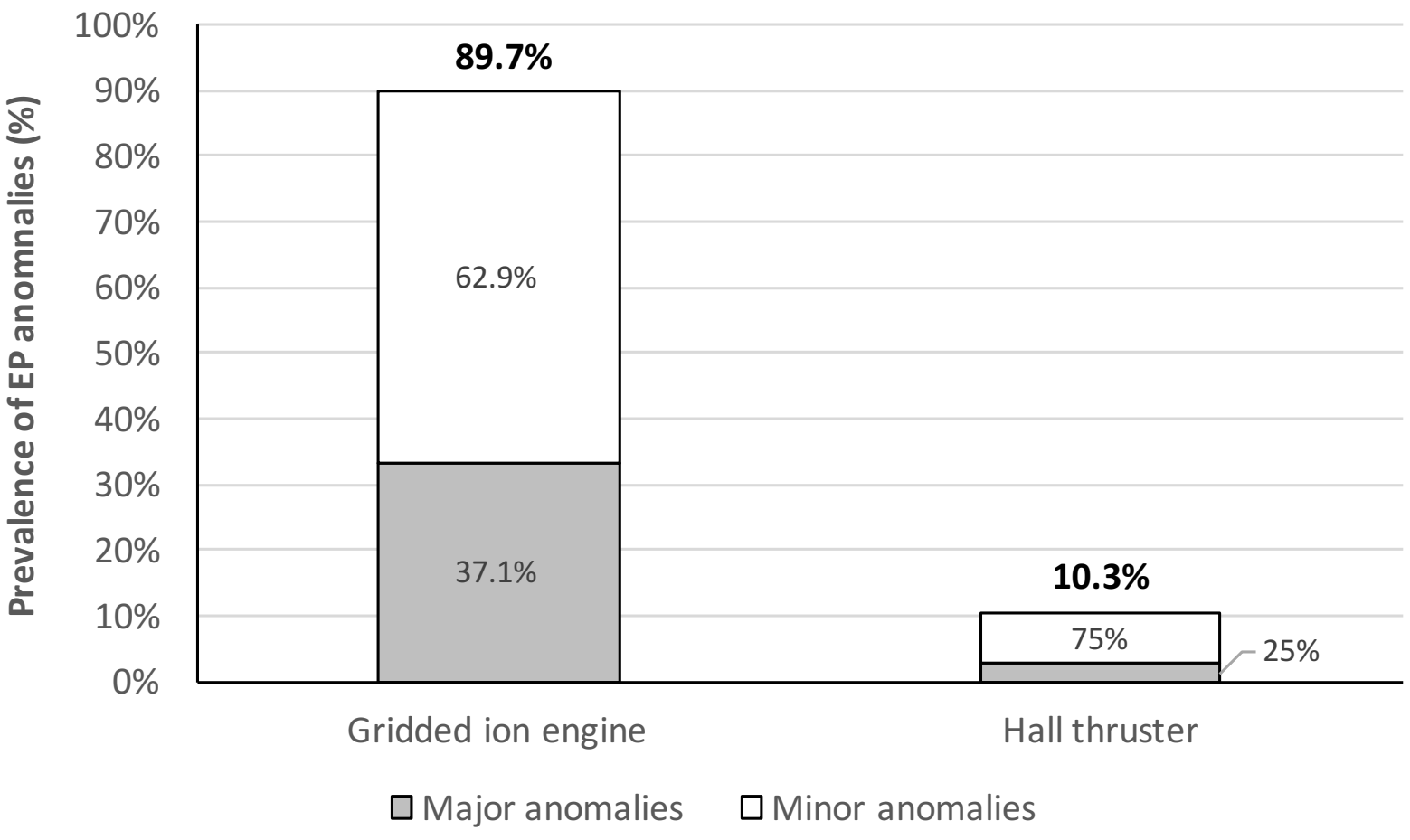

**Figure 7**. Prevalence of EP anomalies by engine type and by severity (n = 39).

When normalized by the number of EP-equipped spacecraft by the number of EP-equipped spacecraft with each type of engine, the previous results are further accentuated, as shown in the right panel in Figure 8. First note that the Hall thrusters are the most prevalent engine type on orbit, equipping 102 spacecraft in our dataset, compared with the 59 gridded ion engine equipped spacecraft. The right panel in Figure 8 shows that gridded ion engines have exhibited an average of 0.59 anomalies per spacecraft so equipped compared with an average of 0.04 anomalies for the Hall thruster equipped spacecraft—a difference of over an order of magnitude for the ion engines.

---

[8] More precisely, the null hypothesis that ion engines and Hall thrusters have the same ratio of minor to major anomalies, cannot be rejected (*p-value* = 0.3).

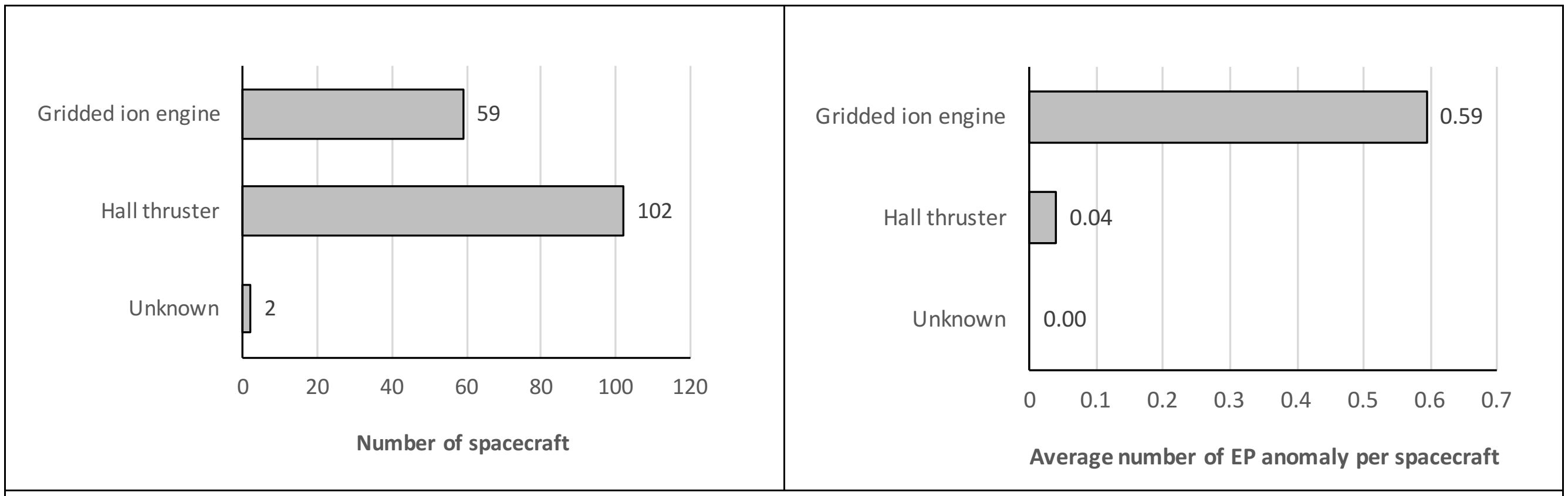

**Figure 8**: Total number of EP-equipped spacecraft by engine type (left panel), and average number of anomalies per spacecraft by engine type (right)

To eliminate the effect of possible *dependent* EP anomalies on the same spacecraft, we considered again only the number of spacecraft that exhibited EP anomalies (not the number of anomalies), and we calculated the risk ratio for the two types of thrusters along with the 90% confidence interval for this statistic.

**Table 5**. Contingency table of spacecraft with and without EP anomalies by engine type

| | **Gridded ion engine** | **Hall thruster** | Total |
|---|---|---|---|
| **Number of spacecraft with EP anomalies** | 19 | 1 | 20 |
| **… without EP anomalies** | 40 | 101 | 141 |
| Total | 59 | 102 | 161* |

* One spacecraft included electro-spray thrusters, and as a result it was not included in this table.

$$P(GIE_anomaly|S/C_GIE) = \frac{19}{59} = 0.322 \qquad (Eq. 4)$$

$$P(HT_anomaly|S/C_HT) = \frac{1}{102} = 0.0098 \qquad (Eq. 5)$$

$$Relative\ Risk, RR = \frac{P(GIE_anomaly|S/C_GIE)}{P(HT_anomaly|S/C_HT)} = 32.8 \qquad (Eq. 6)$$

The 90% confidence interval around the relative risk *RR* is given by first calculating it around the

ln($RR$), which according to statistical theory (see for example Rothman et al., 2008) is approximately normally distributed, then by exponentiating:

$$\ln(RR) \pm 1.6448 \cdot \sqrt{\frac{\frac{40}{19}}{59} + \frac{\frac{101}{1}}{102}} = [1.82; 5.15] \qquad (Eq. 7)$$

$$\Rightarrow$$

$$90\%\ CI \quad RR = [6.2; 172.4] \qquad (Eq. 8)$$

Equation 6 shows that the risk of a spacecraft experiencing an EP anomaly is 32 times higher for gridded ion engine equipped spacecraft that for a Hall thruster equipped spacecraft. Because of the relatively small size of the sample, we compute the 90% confidence interval for this risk ratio, as shown in Eq. 7 and 8. Since the null value of the risk ratio ($RR_0$ = 1) is not included in the confidence interval, we reject the null hypothesis and assert that there is **strong evidence against the statement that spacecraft equipped with gridded ion engines exhibit the same proportion of EP anomalies as those equipped with Hall thrusters** (*p-value* = 0.0002)**. Said more casually, spacecraft with gridded ion engines experience a statistically significant (much) higher risk of anomalies than those equipped with Hall thrusters**.

Two concerns can be raised with regard to these calculations:

1. First, since the previous subsection showed that EP anomalies and orbit type are dependent, these results (Eq. 6 and 8) might be confounded by orbit type. This is a valid concern, and to address it, we stratified the data by orbit type. Only GEO had sufficient data point to allow a similar analysis. The results showed a risk of anomaly for gridded ion engines of 27% in GEO, instead of 32% for aggregated data in Eq. 4 (all orbits), and a risk difference between ion engines and Hall thrusters in GEO of 27 percentage points instead of the 31 percentage points for the aggregated data. As a result, we believe orbit type has not confounded the results in Eq. 4–8, or if it has, the bias is minimal and does not change the conclusion.

2. By accounting for spacecraft that exhibited anomalies, not the number of anomalies, we risked underestimating the differences between the two types of engines and bias the results toward the null hypothesis. This is a valid concern. The reason for our choice was that a spacecraft is typically equipped with four thrusters, and should a single spacecraft exhibit multiple anomalies, these are likely to be dependent. To mitigate that possibility, we examined here the number of spacecraft (not the number of anomalies). The results should indeed be viewed as conservative and they under-estimate the difference in likelihood of anomalies in gridded ion engines in general (not spacecraft equipped with such engines) and Hall thrusters (compare with Figure 8, right panel).

The results for the Mean Time To Anomaly by engine type and by severity are provided in Figure 9. The results show a statistically significant difference between the $MTTA_{minor}$ of gridded ion engines and Hall thrusters: the latter exhibit a very short MTTA of 116 days compared with the 845 days for ion engines. This difference is statistically significant (*p-value* = 0.00005) and can be casually remembered as follows: **Hall thruster exhibit minor anomalies very early on orbit—in the first few months, which is indicative of infant anomalies, and thus would benefit from better ground testing, improved quality screening, and dedicated acceptance procedures—and then they get their act together after that and perform almost flawlessly**[9]. Gridded ion engines on the other hand exhibit a widely varying time to anomaly, with an $MTTA_{minor}$ of over two years. The individual times to minor anomaly for the **gridded ion thrusters** are bimodal and **exhibit both infant and wear-out failures**, 9 minor anomalies in the first year after launch, and 7 after 4 years, and a few in between, and as a result, **gridded ion thrusters would benefit from a reliability growth program that addresses both types of problems, the infant anomalies and the wear-out failures**. These statistical differences are likely indicative of different failure modes or mechanisms for these two types of engines.

The larger $MTTA_{major}$ for both engine types compared with their respective $MTTA_{minor}$, e.g., 1443 days for major anomalies in gridded ion engines compared with their 845 days to minor anomalies, shows a progression toward more serious failures, which is indicative of wear-out, fatigue, and/or erosion mechanisms. However, these differences are not statistically significant (probably due to the small sample size), and consequently judgment as to their robustness should be cautiously withheld.

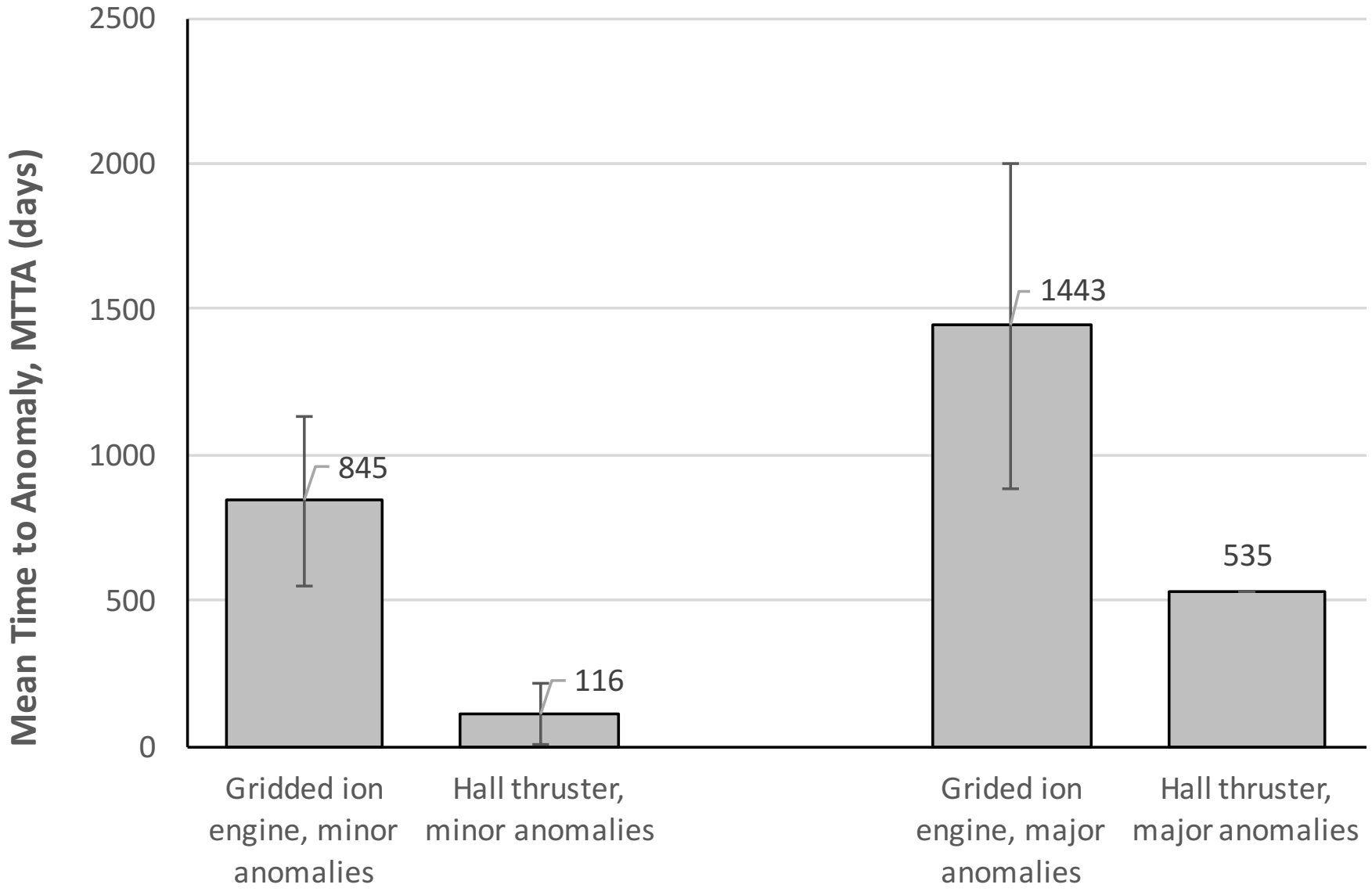

**Figure 9**. Mean Time To (minor/major) Anomaly by engine type, along with the 90% confidence intervals (major anomaly not shown for the Hall thruster since only one such event was recorded)

[9] No other minor anomalies occurred after that, and only one major anomaly occurred after 535 days on orbit (about a year and a half). The track record is flawless after this.

## 4. Comparative analysis: EP versus chemical propulsion failure rates in GEO

In this section, we eliminate the confounding effect of orbit type on anomalies by restricting our analysis to GEO. This is a particularly important *real estate* on orbit since over half of the \$200+ billion space industry is based in or related to systems in GEO (Wade et al., 2015). We examine the failure rates of EP over the entire time period in our sample (1997–2015), then for reason discussed later, we redo the calculations over two time periods, 1997–2004 and 2005–2015.

One important question for satellite designers, satellite operators, and insurers is, how does the reliability of EP compare with that of chemical propulsion? The purpose of this question is to assess and account for the risk differential, if any, among other criteria when deciding between the adoption of EP or chemical propulsion, or when promoting one or the other. It is important to make a reliability- and risk-informed decision in this regard. To help in this process, we also examine in this section the failure rates of chemical propulsion in GEO satellites over the same two time periods. We then compare both modes of propulsion and conclude whether one outperforms the other or not in terms of reliability.

### 4.1 EP anomaly rates in GEO

Figure 10 shows the net cumulative number of EP-equipped spacecraft in GEO since 1997 (active in a given year). This cumulative count subtracts the number of spacecraft retired in a given year (net). Also shown are the number of EP anomalies in a given year. The salient features in this figure are the following: (1) a significant and sustained growth of EP-equipped spacecraft between 1997 and 2015, with a compounded annual growth rate (CAGR) of about 16%. It is interesting to note that the net addition of EP-equipped spacecraft to GEO has varied between 4 and 12 per year, with a median of 7 and an interquartile range of 6.5 (4.25 to 10.75 EP-equipped spacecraft per year to GEO for the first and third quartile); (2) a dramatic peak of EP anomalies in 2000, followed by a secondary peak in 2003; and (3) the absence of EP anomalies in GEO since 2011. It should be pointed out that since no Hall thruster anomalies occurred in GEO, the results in Figure 10 show gridded ion engine anomaly counts.

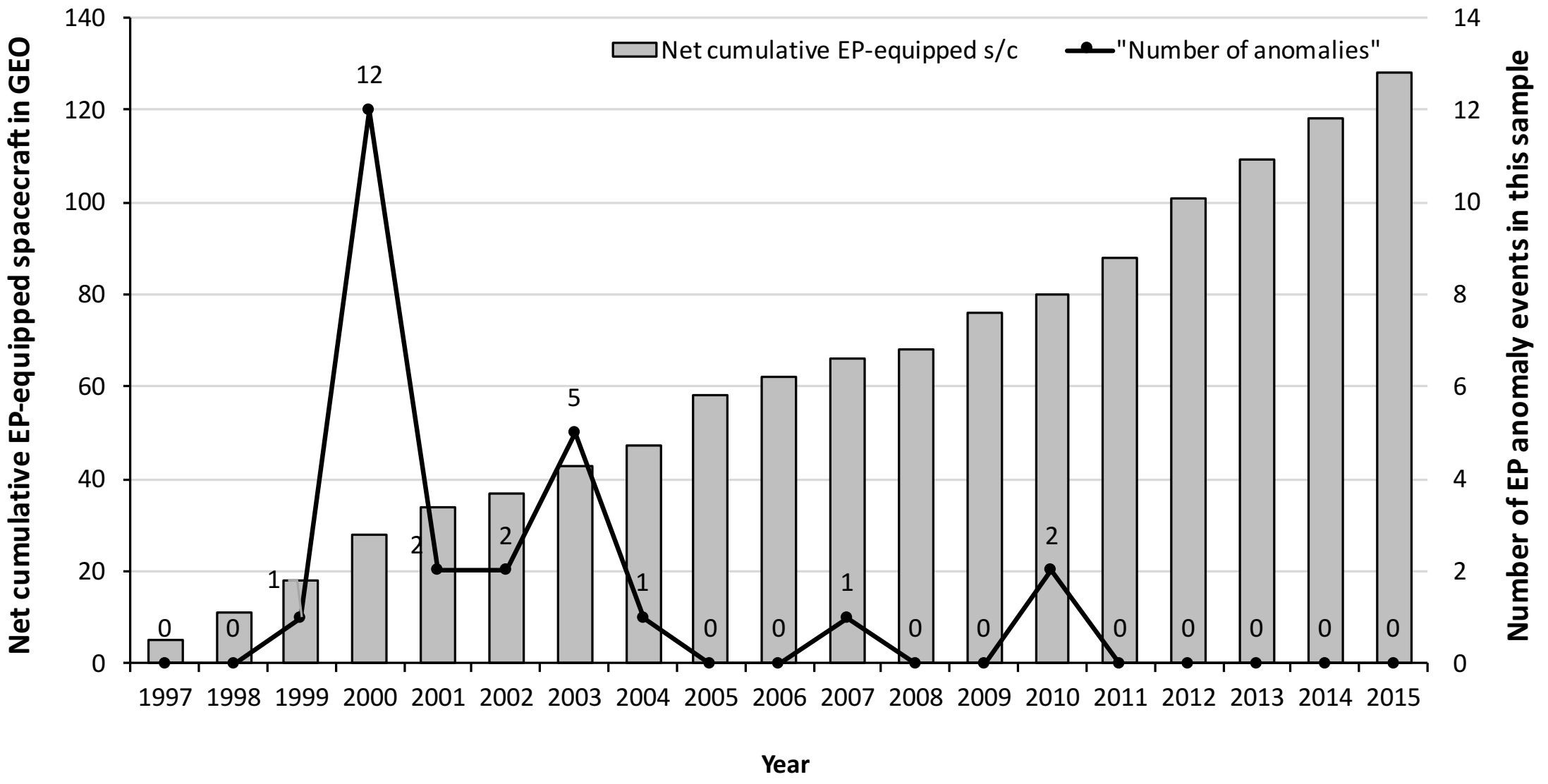

**Figure 10**. Net cumulative EP-equipped spacecraft in GEO and the number of EP anomalies in a given year since 1997 (gridded ion engines only; Hall thrusters exhibited no anomalies in GEO during this time period)

The 3-year rolling average anomalies per spacecraft in Figure 11 shows the patterns more clearly (albeit with a lead time for the peak average anomaly per spacecraft in 1999 instead of 2000 because of rolling average calculations). The two important features in Figure 11 are the steady decrease in average anomalies per spacecraft between 2000 and 2004, and the significantly different (and smaller) average anomalies per spacecraft pre- and post-2005.

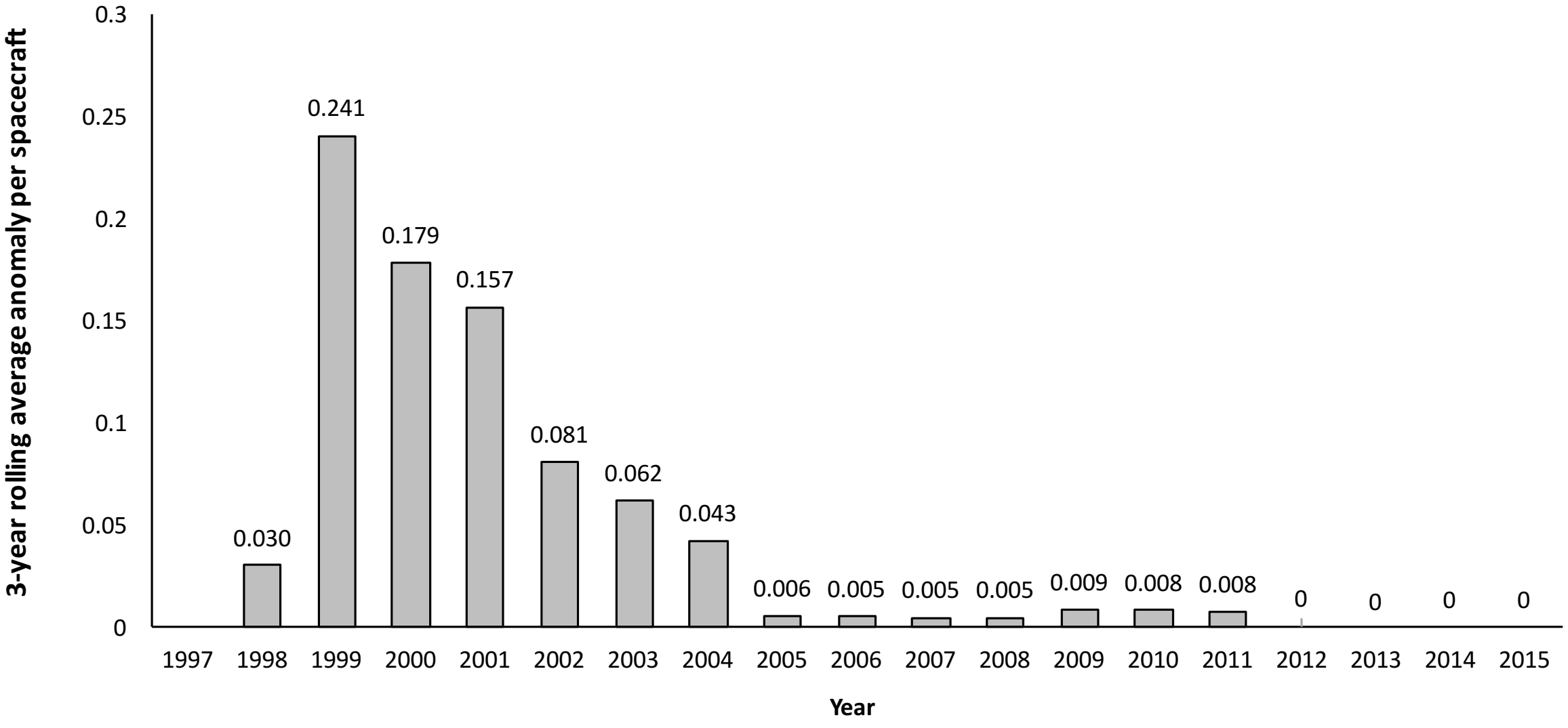

**Figure 11**. Three year rolling average number of EP anomalies per spacecraft in GEO

To calculate actual rates of anomalies, we first had to assess the number of spacecraft-day each EP-equipped spacecraft contributed to our dataset. This requires some explanation. In epidemiology, the term *rate* is formally defined as "the occurrence of new cases of disease that arise during person-time of observation" in a given population (Aschengrau and Seage, 2008), that is, time is intrinsic to the denominator in rate calculations[10]. For our purpose, we calculate **spacecraft-time** instead of person-time, which is accrued only when the spacecraft is launched and until it is retired or completely fails on orbit. Figure 11 does not reflect rates since spacecraft in a given year can be launched during any month of the year (not just in January), and thus do not contribute a full year to the denominator of rate calculation, and some spacecraft can be retired before the end of the year yet they contribute some spacecraft-time in a given year to the denominator.

To mitigate these issues, we tracked each EP-equipped spacecraft in GEO and calculated its spacecraft-time contribution to our sample (from launch until retirement, or Class I failure event, or until the end of our observation window of December 31, 2015). Overall, EP-equipped spacecraft in GEO contributed 399,384 spacecraft-days between 1997 and 2015 (1093 years or about 9.6 million hours). The split between 1997–2004 and 2005–2015 is shown in Table 6.

**Table 6**. Spacecraft-time in GEO for EP-equipped spacecraft

| | **1997–2004** | **2005–2015** |
|---|---|---|
| Spacecraft-days (years) | 71,133 (194.7) | 328,251 (898.7) |

With these numbers, we then calculate anomaly rates over these two time periods. The results are shown in Figure 12. The y-axis is kept in spacecraft-day instead of the more convenient spacecraft-year to avoid on the one hand possible confusion with the results of the rolling average (per year) in Figure 11, and to indicate on the other hand that the level of resolution in the underlying calculations was indeed a spacecraft-day, not the more crude spacecraft-year.

[10] The term *failure rate* in reliability analysis is a conditional probability density function, f(t)/(1-F(t)). Our calculations are the statistical implementation of this probabilistic definition.

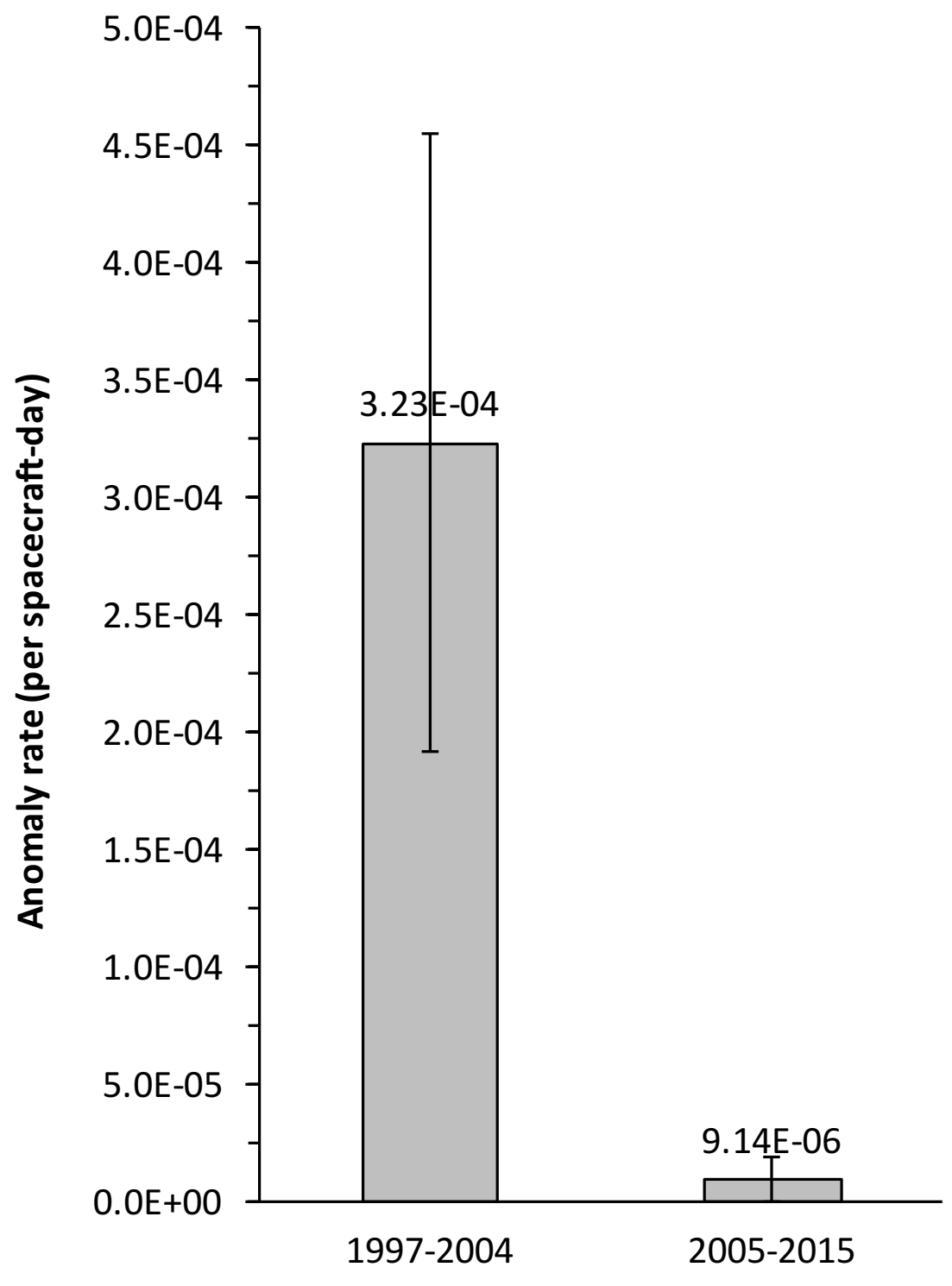

**Figure 12**. Anomaly rates of EP-equipped spacecraft in GEO launched between January 1, 1997 and December 31, 2015. (gridded ion engines only; Hall thrusters exhibited no anomalies in GEO during these time periods)

**Figure 12 displays an important result, and it confirms that the EP anomaly rates pre- and post-2005 are indeed non-homogeneous and they exhibit statistically significant differences during these two time periods**. Recall these are only for gridded ion engines since Hall thrusters exhibited no anomalies during these time periods. The anomaly rate pre-2005 was $3.23 \cdot 10^{-4}$ per spacecraft-day (or about 0.118 per spacecraft-year), and it drops a staggering 35 folds post-2005 to $9.14 \cdot 10^{-6}$ per spacecraft-day (or about 0.003 per spacecraft-year).

It is interesting to reflect on this result, and in jointly considering Figures 10 and 12, **one is led to surmise that perhaps a typical reaction time in the space industry for identifying and fixing a problem with a subsystem—here gridded ion engines—is about four to five years**. There are organizational and structural issues in the space industry that lend credence to such a hypothesis. This of course deserves more careful consideration and is left as a fruitful venue for future work.

The EP anomaly rate results are further broken down by severity in Figure 13. No statistically significant difference existed in minor versus major anomaly rates pre-2005, and while the minor anomalies appear to have been completely eliminated post-2005, a negligible major anomaly rate remained post-2005.

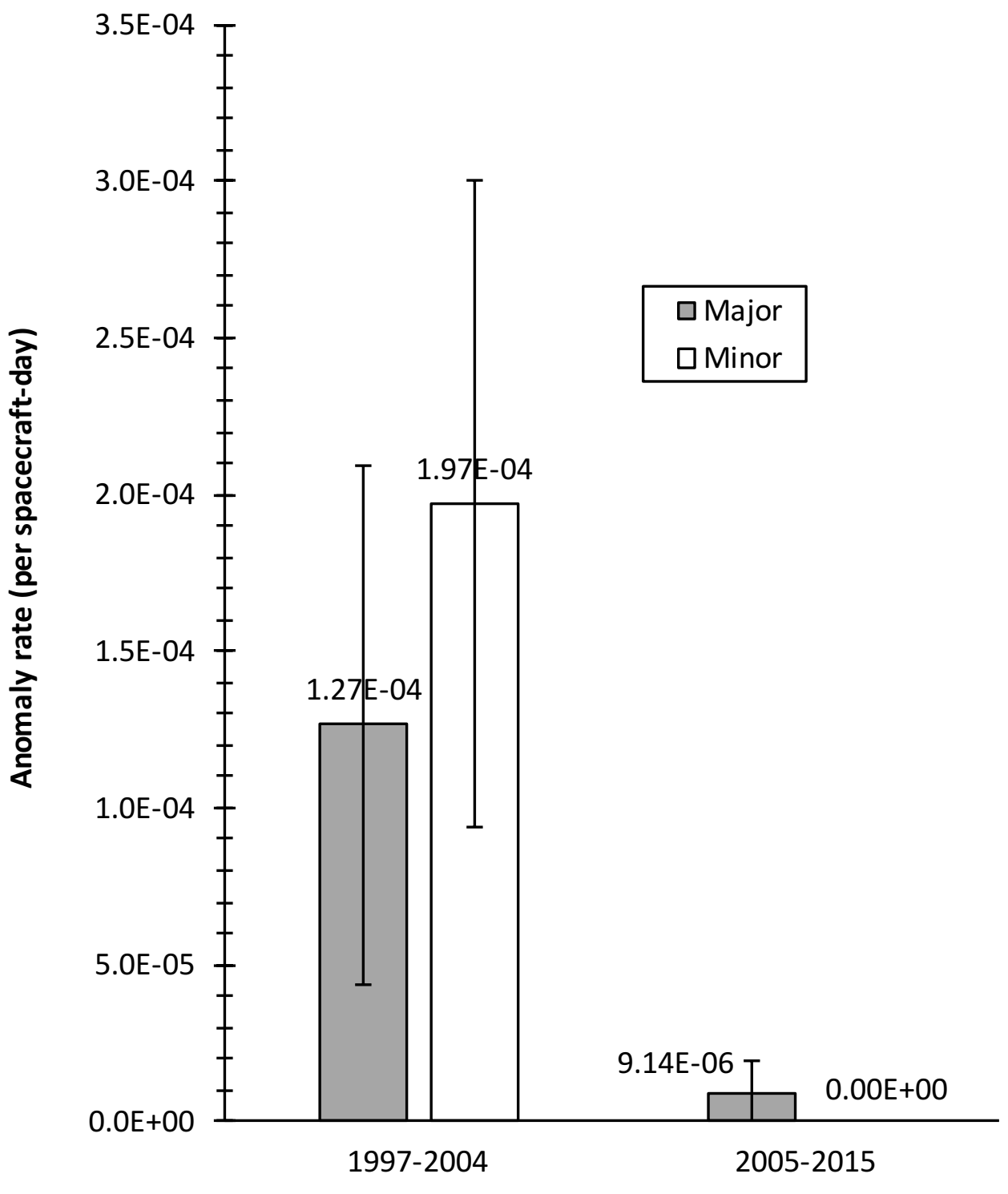

**Figure 13**. Anomaly rates of EP-equipped spacecraft in GEO by severity (gridded ion engines only; Hall thrusters exhibited no anomalies in GEO during this time period)

The important remaining question is how do these EP anomaly rates compare with those of chemical propulsion in the same orbit and over the same time period? To address this question, we first collected all the chemical propulsion anomaly and failure data for GEO spacecraft launched between January 1, 1997 and December 31, 2015. We then conducted similar analyses to the ones reported in this subsection. The results are provided next.

### 4.2 Chemical propulsion anomaly and failure rates in GEO

Figure 14 shows the net cumulative number of CP-equipped spacecraft in GEO since 1997 (active in a given year). Unlike our decision to exclude electro-thermal devises from the EP dataset, we placed no restriction on the CP data collected (although the majority of the subsystems were bi/monopropellant thrusters). The figure shows a significant and sustained net growth of chemical propulsion-equipped spacecraft in GEO, hereafter referred to as CP-equipped spacecraft, with a compounded annual growth rate (CAGR) of about 15%, very similar to the growth rate of EP-equipped spacecraft in GEO (about 16%). It is interesting to note that the net launches of CP-equipped spacecraft to GEO has varied between 10 and 34 per year, with a median of 20 and an interquartile range of 7 (16 to 23 CP-equipped spacecraft per year to GEO for the first and third quartile). This along with the data for EP-equipped spacecraft provides a good indication of the size of the GEO market per year, with a median of 26 spacecraft per year and an interquartile range of about 11 (21 and 32 for the first and third quartile respectively).

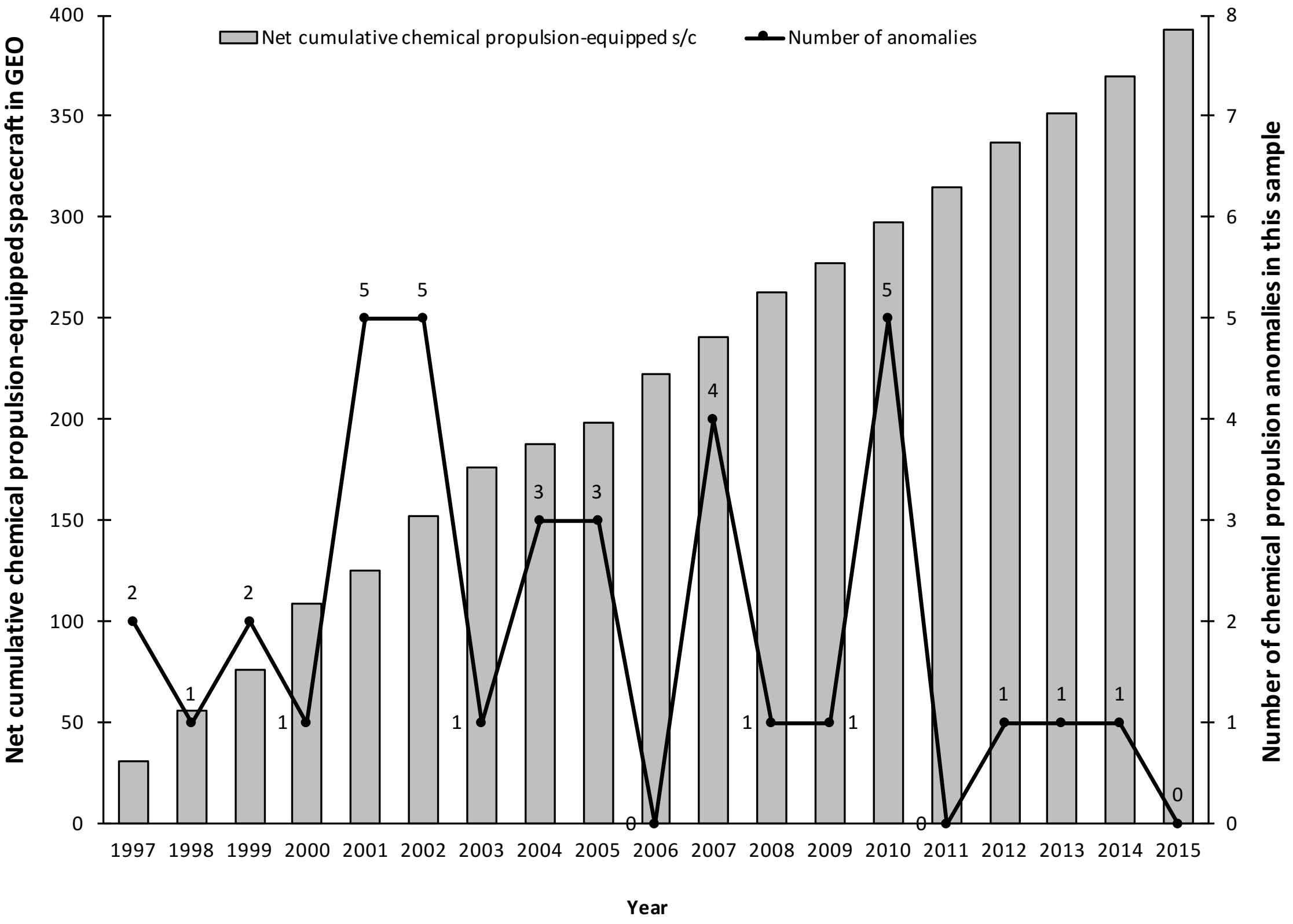

**Figure 14**. Net cumulative chemical propulsion-equipped spacecraft in GEO and the number of chemical anomalies in a given year since 1997

A total of 37 failure events occurred during this time period in the chemical propulsion subsystem of GEO spacecraft. One particularly important result not visible in Figure 14 is that 6 of these events were Class I failures, that is, they led to the complete loss of the spacecraft. A rough back-of-the-envelope calculation suggests that for a typical cost to Initial Operational Capability (IOC) of a GEO spacecraft of about $250m, and given that the Mean Time To (Class I) failure was about 1000 days, **the pro-rated losses due to the chemical propulsion subsystem are probably about $1.2 billion between 1997 and 2015 (only in GEO)**.

Table 7 provides the causes reported for the chemical propulsion anomalies and failures (Class I and Class II). In some instances, no details are available either because the details were not disclosed by the operator, or because the investigation has not converged on a specific cause beyond "thruster malfunction" (this is a reflection of the extent of satellite telemetry and health monitoring points).

**Table 7**. Reported causes of chemical propulsion anomalies and failures (Class I and Class II)

| **Chemical Propulsion: reported causes of class I failures* (total loss)** | **Chemical propulsion: reported causes of Class II anomalies (major anomalies)** |
|---|---|
| Leak in fuel cooling system compartment caused the failure and total loss of the satellite | Problem with overheating/leaking thrusters due to degradation of sealing |
| Satellite has reportedly depressurized. Believed to have been caused by a leak (seal failure) in the propellant tank. Total loss of the satellite | Believed to be suffering from leaking thrusters. May have leaked into the interior of the satellite causing contamination (reported in three different satellites) |
| Faster than expected fuel usage (details not identified/disclosed). Satellite Moved into a graveyard orbit. Effective retirement/loss of the satellite (reported in other instances) | Leak in its helium pressurization system. Originally it was expected that this would reduce satellite life by five years but now only 20 months is to be lost |
| The satellite had two different propellant tanks, which resulted in an unequal flow of fuel. A design fault meant that this had not been correctly accounted for. The resulting imbalance made the spacecraft tilt, and as a result, all the propellant was depleted in order to stabilize the satellite. The satellite was effectively retired (total loss) | Fuel system anomalies were confirmed and useful life was estimated as being reduced to four years remaining. Details not disclosed<br><br>Thruster malfunction during North-South inclination maintenance manoeuver (no details available) |
| | A thruster experienced an anomaly and was withdrawn from service (no details provided). Another thruster had an anomaly and was withdrawn from operation. The satellite is equipped with a total of 12 thrusters and is currently operating using a combination of the other 10 thrusters. This workaround requires more frequent manoeuvers, which are less efficient and therefore result in accelerated fuel use<br><br>An investigation of the thruster anomalies including the development of additional workarounds for long term operations is ongoing |

* Time to failures roughly bi-modally distributed, with one Mean Time To Failure or MTTF = 42 days and another MTTF = 3871 days

We skip the rolling average results for the CP-equipped spacecraft in GEO, and provide the more informative anomaly rate by severity in Figure 15.

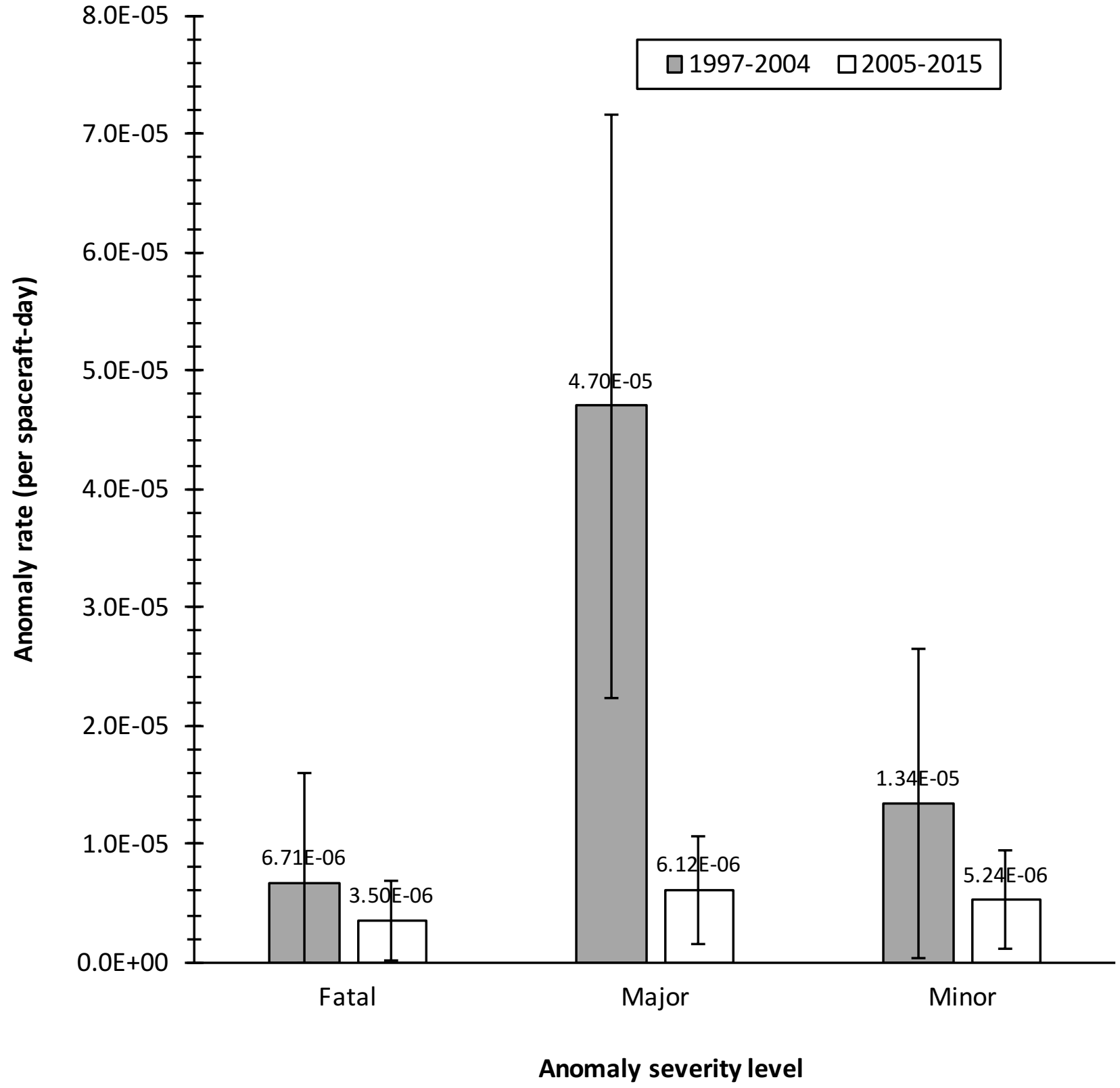

**Figure 15**. Anomaly rates of chemical propulsion-equipped spacecraft in GEO by severity

Figure 15 shows that for the fatal and minor anomalies, no statistically significant decrease or compression of the anomaly rates occurred pre- and post-2005. However, a statistically meaningful decrease occurred for the major anomaly rate between these two time periods. It is worth venturing into the realm of speculation for a brief moment at this point and suggest that **this finding perhaps reflects on the one hand that better design, testing, and quality control procedures have eliminated failure modes that led to major anomalies in the chemical propulsion subsystem, but that other failure modes that lead to minor and fatal failures still exist and continue to elude designers and testers**. This hypothesis deserves more careful examination and is an interesting direction for future work.

### 4.3. Comparative analysis: anomaly rates of EP versus chemical propulsion in GEO

We begin by comparing overall anomaly rates of EP versus CP, irrespective of severity. Two important results are shown in Figure 16:

1. First, **EP was significantly more troublesome pre-2005 than chemical propulsion in terms of frequency of overall anomalies. More specifically, EP was about five times more likely to exhibit an anomaly than a chemical propulsion subsystem in GEO**;

2. Second, the difference in overall anomaly rates all but vanished post-2005, and at present EP and CP in GEO exhibit no statistically significant differences in terms of their failure rates (notice the highly overlapping 90% confidence intervals post-2005; the slight advantage for EP is not statistically significant). We can thus conclude that there is no evidence **post-2005** to reject the statement that **EP and chemical propulsion exhibit the same overall anomaly rate**.

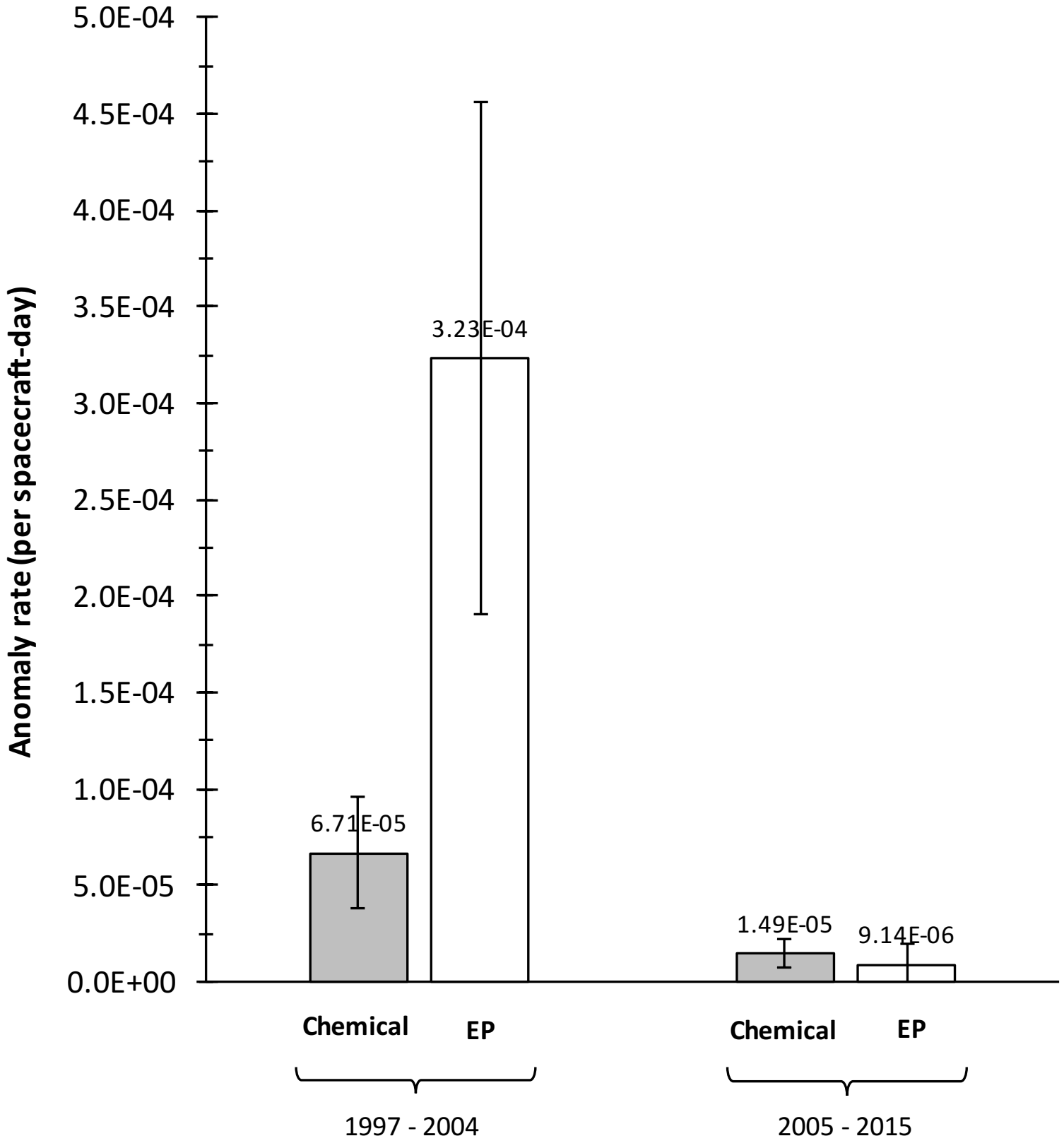

**Figure 16.** Anomaly rates of EP versus chemical propulsion-equipped spacecraft in GEO (color code in this figure simply refers to the type of propulsion system)

The comparative analysis by severity of the failure event (Figures 17 and 18) provides more nuances to the previous conclusion and adds one particularly important result to the mix. First note that pre-2005, EP was significantly more troublesome than chemical propulsion in terms of frequency of major and minor anomalies, as seen in Figure 17: over an order of magnitude more frequent minor anomalies, and about 3 times more major anomalies, the latter not statistically significant.

A more nuanced conclusion is therefore warranted: **pre-2005 in GEO, EP exhibited a significantly higher frequency of minor and major anomalies than chemical propulsion.**

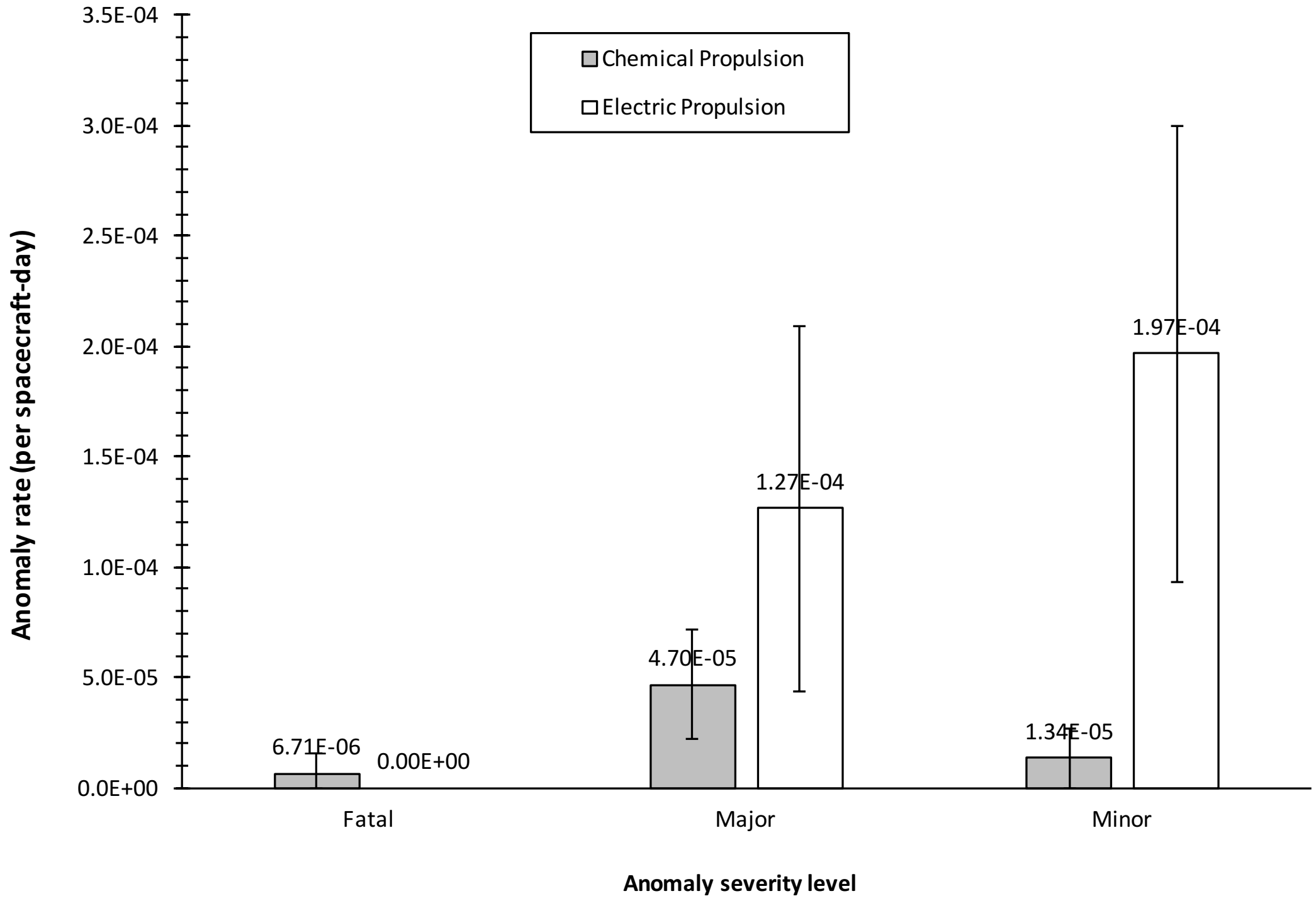

**Figure 17.** GEO pre-2005: comparative analysis of anomaly rates by severity of EP versus chemical propulsion-equipped spacecraft

The situation post-2005 has completely changed, as shown in Figure 18.

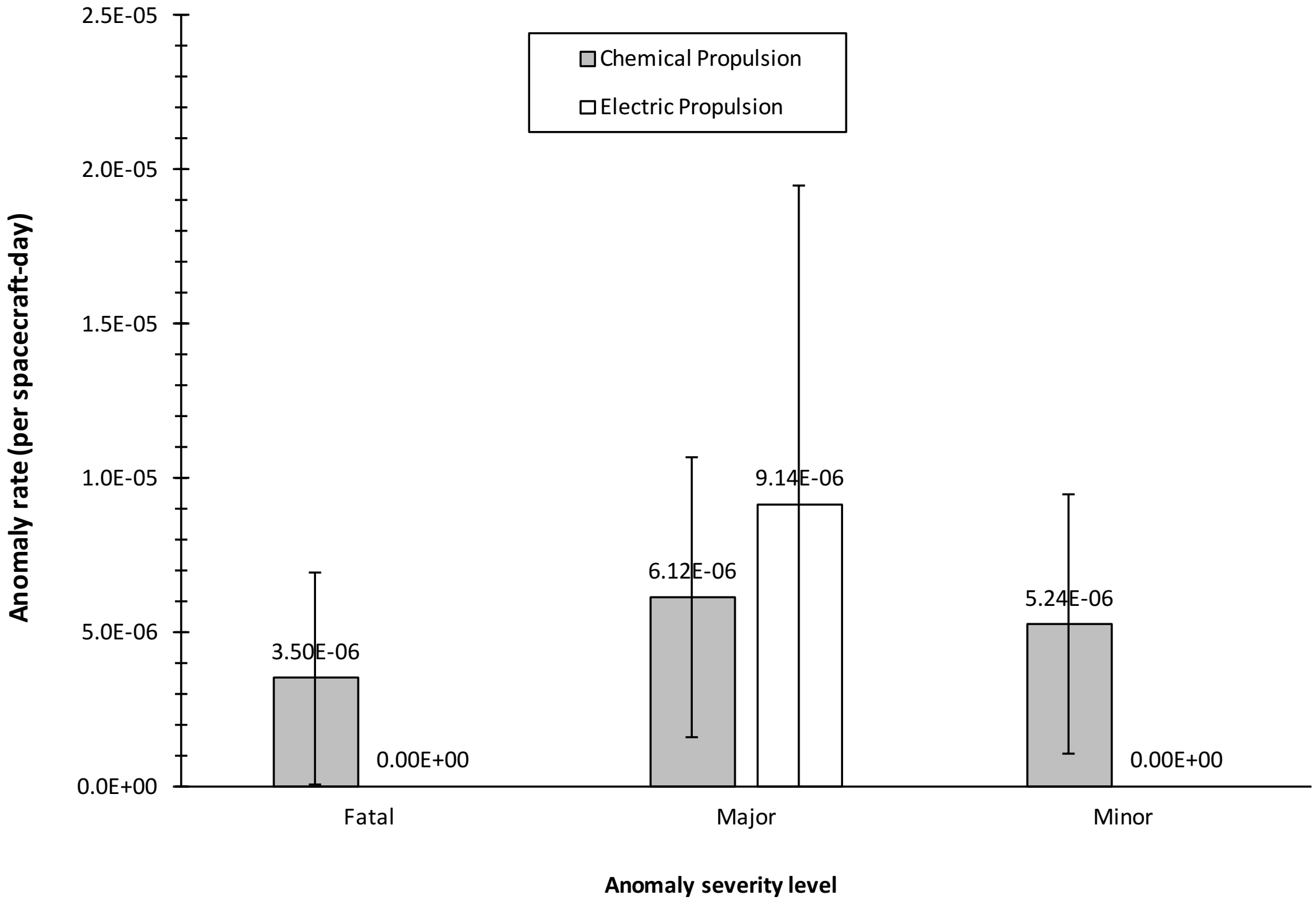

**Figure 18.** GEO post-2005: comparative analysis of anomaly rates by severity of EP versus chemical propulsion-equipped spacecraft

**While all the failure rates have been compressed compared with their pre-2005 values, the most aggressive improvements occurred with electric propulsion. EP exhibited during this time period smaller (statistically significant) rates for the fatal and minor anomalies than chemical propulsion**. However, the major anomaly rates for EP and CP are not statistically different. Consequently, it is fair to conclude that **post-2005 in GEO, EP has outperformed chemical propulsion in terms of reliability when both frequency (rate) and severity of anomalies are examined.**

It is worth repeating a point noted earlier in this section that no Hall thruster anomalies occurred in GEO during the 1997–2015 time-period[11], and therefore the results in this section concern gridded ion engines. The consequences should not be ignored, and it is important to state them explicitly:

[11] Should data quality be suspected in these results (e.g., Hall thruster anomalies on GEO Russian not reported), it is useful to note that some 45 Western spacecraft in GEO use Hall thrusters (compared with 55 with gridded ion engines), and these results still hold true when the analysis is restricted to Western spacecraft.

**Pre-2005 in GEO, Hall thrusters have robustly outperformed chemical propulsion and gridded ion engines. Post-2005, significant improvements have occurred with gridded ion engines, and their minor anomaly rate has dropped to 0 but not their major anomaly rate. The risk difference between gridded ion engines and Hall thrusters is therefore confined to major anomalies and is about 0.33% per year. In short, post-2005 both EP technologies outperform chemical propulsion in terms of reliability; and Hall thrusters maintain a small but shrinking reliability advantage over gridded ion engines.**

### 4.4. Caveat: the role of XIPS-13 and its irrelevance for future decisions regarding the choice of propulsion system[12]

It is useful to acknowledge, for the accuracy of the record, that six (hybrid) spacecraft in GEO equipped with the gridded ion engine XIPS-13 had their primary EP thrusters experience malfunction or fail on orbit[13], and either the secondary failed as well or the satellite manufacturers recommended to the operators not use the redundant EP thrusters (because of the likelihood of a similar/dependent failures). As a result, the satellites switched to the chemical propulsion subsystem instead. Significant losses were incurred because of these EP failures, mainly in the form of reduction in service life of the satellite (i.e., the losses were in the form of revenues forfeited). The satellites however continued to operate for several years after this EP shutdown.

A question and debate arose whether these EP failures should be classified as Class I or Class II failures[14]. The SpaceTrak database defines Class I as a failure leading to the complete loss of the spacecraft; since these six failures did not lead to such outcome—the satellites remained operational for several years after the EP shutdown—the database classified these events as Class II failures. Class II consists of failures that lead to significant and permanent damage, but not total destruction of the satellite.

It is fair to examine more carefully this classification though: should this EP shutdown be considered a Class I failure or not? Unlike the Class I failures due to other subsystems, including chemical propulsion (e.g., depressurization of propellant tank, or leak within the satellite), which lead to the immediate death and complete loss of the satellite, the EP failures and shutdown in these instances did not lead to such

---

[12] We are grateful for an anonymous reviewer who suggested that we add this subsection for clarity.

[13] The satellites are Galaxy VIII-i, PAS-5, PAS-6B, SatMex-5, Galaxy-10R, and Galaxy IVR.

[14] Recall the definitions of these two classes of failures:

- Class II: major non-repairable failure that affects the operation of a satellite or its subsystems on a permanent basis. This effectively means major/significant losses, but not total immediate loss of the satellite;
- Class I: subsystem failure causing satellite retirements. This effectively means immediate and total loss of the satellite.

dramatic consequences. The definition of the Class II failures includes these EP-shutdown instances.

The argument however was made that, "had chemical propulsion not been available in conjunction with the EP thrusters (XIPS-13), then these satellite would have been lost", thus leading to a hypothetical Class I failures. Although the merits of this arguments should be acknowledged, it is difficult to deal with such hypotheticals, first, because the facts are clear: these six satellites were not lost following the EP shutdown, and they remained in operation for several years afterward. So factually, these instances cannot be classified as Class I failures (consistent with the definition of this class of failure events). Second, just a hypothetical alternative was formulated, "had chemical propulsion not been present…", another equally valid hypothetical can be formulated: "had the satellite manufacturer and/or operator not made this particular choice of diversity of redundancy of the propulsion system (EP and CP), then another form of redundancy would have been adopted. In which case, there is no telling whether the satellite would have completely failed after the main EP shutdown.

Both hypotheticals are equally valid, unresolvable, and pointless at this point. We chose to remain consistent with the facts and the classification adopted by SpaceTrak. These considerations, while useful for the past track record of the XIPS-13, are of no relevance for future decisions regarding the choices of space propulsion since this particular thruster is no longer in production.

### 4.5. Limitation

Is it fair to compare the anomaly rates of EP and CP in GEO, as we did in the previous subsection? One final limitation of the comparative analysis is worth acknowledging. The functions fulfilled to date by EP and CP are overlapping but not completely identical. For example, while both propulsion subsystems have been used to perform station-keeping and orbit maintenance, chemical propulsion has been the main workhorse for orbit raising. In only three cases we are aware of was electric propulsion used for orbit raising to GEO. Consequently, the risk of anomaly during this phase is almost exclusively carried by chemical propulsion. To account for this difference in usage/function performed between EP and CP, one would need to further stratify the anomaly data by phase, namely orbit raising and station-keeping. However, given the quasi-absence of experience of EP orbit raising to date, no such analysis can yet be meaningfully conducted (this would probably be possible in a few years from now, when more experience/data with EP orbit raising is gained). It is important therefore to keep in mind that the results in 4.3 are blind to the possibility of such a confounding effect.

Having recognized this limitation, we believe the resulting bias, should it exist, is small as the mean time to major anomaly (MTTA) for CP is 992 days, significantly removed from the orbit raising phase, and that one of the modes for the mean time to failure (Class I MTTF) is 3871 days. No similar data exists for the EP involved in orbit raising. The possibility of a bias exists due to the other mode of mean time to failure of 42 days (see Table 7), which is within the vicinity but on the outer edge of the orbit raising phase.

## 5. Conclusion

Prior to the 1990's, EP occupied a small niche market in the space industry. Several factors discussed earlier in this work conspired to delay its development and adoption, and kept it stuck in the slow maturation lane. At present, the situation has significantly changed, and this hundred-year old idea of electric propulsion appears to be having its day and turning the tables on chemical propulsion in space. It is thought-provoking to consider that perhaps the bell has begun to toll for spacecraft chemical propulsion. In the next decade, the market share of spacecraft chemical propulsion will continue to erode and is likely to become confined to a small niche market of missions with high thrust and high maneuverability requirements.

Why this forecast?

It was noted earlier that despite the advantages that EP provides (e.g., higher specific impulse and significantly lower mass than chemical propulsion), electric propulsion will not be broadly adopted until two fundamental analyses have been conducted and questions addressed: a value analysis on the one hand integrating the various benefits and costs of EP, and on the other hand a reliability/risk analysis, and benchmarking both against spacecraft with chemical propulsion. The present work addressed the second question. Given the high cost of access to space and the quasi-unavailability of on orbit maintenance to compensate for subpar (hardware) reliability, the space community is understandably risk averse, and this risk aversion explains in part the earlier slow uptake of EP. Satellite operators, manufacturers, and insurers make reliability- and risk-informed decisions regarding the adoption and insurance of particular technologies on board spacecraft. One important question for these stakeholders is, how does the reliability of EP compare with that of chemical propulsion? The purpose of this question is to assess and account for the risk differential, if any, among other criteria when deciding between the adoption of EP or chemical propulsion, or when promoting one or the other. This work provided answers to this question. More specifically, it was shown that[15]:

1. Post-2005, EP has outperformed chemical propulsion in terms of reliability, when both the frequency and severity of anomalies are examined—except for Class II major anomaly rates for which EP and CP exhibit no statistically significant differences (i.e., the statement that they are the same cannot be rejected);
2. Hall thrusters have robustly outperformed chemical propulsion in terms of reliability;
3. Hall thrusters maintain a small but shrinking reliability advantage over gridded ion engines.

Other results were also provided, for example the differentials in MTTA of minor and major anomalies for gridded ion engines and Hall thrusters. It was shown for example that:

[15] It is important to keep in mind that electro-thermal devices were excluded from our analysis (for reasons discussed in 2.2), and as such, the results that follow are confined to electromagnetic and electrostatic propulsive devices.

4. Hall thruster exhibit minor anomalies very early on orbit—in the first few months, which is indicative of infant anomalies, and thus would benefit from better ground testing, improved quality screening, and dedicated acceptance procedures;
5. Gridded ion thrusters exhibit both infant and wear-out failures, and as a result, gridded ion thrusters would benefit from a reliability growth program that addresses both types of problems, the infant anomalies and the wear-out failures.

It was surmised that these statistical differences are likely indicative of different failure modes or mechanisms for these two types of engines. These and other questions—the physics of failure of EP anomalies, and the hypotheses that differences in duty cycle and/or the space environment mediate between orbit type and EP anomalies (frequency and onset)—were left as fruitful venues for future work.

## Appendix: Weibull probability plots

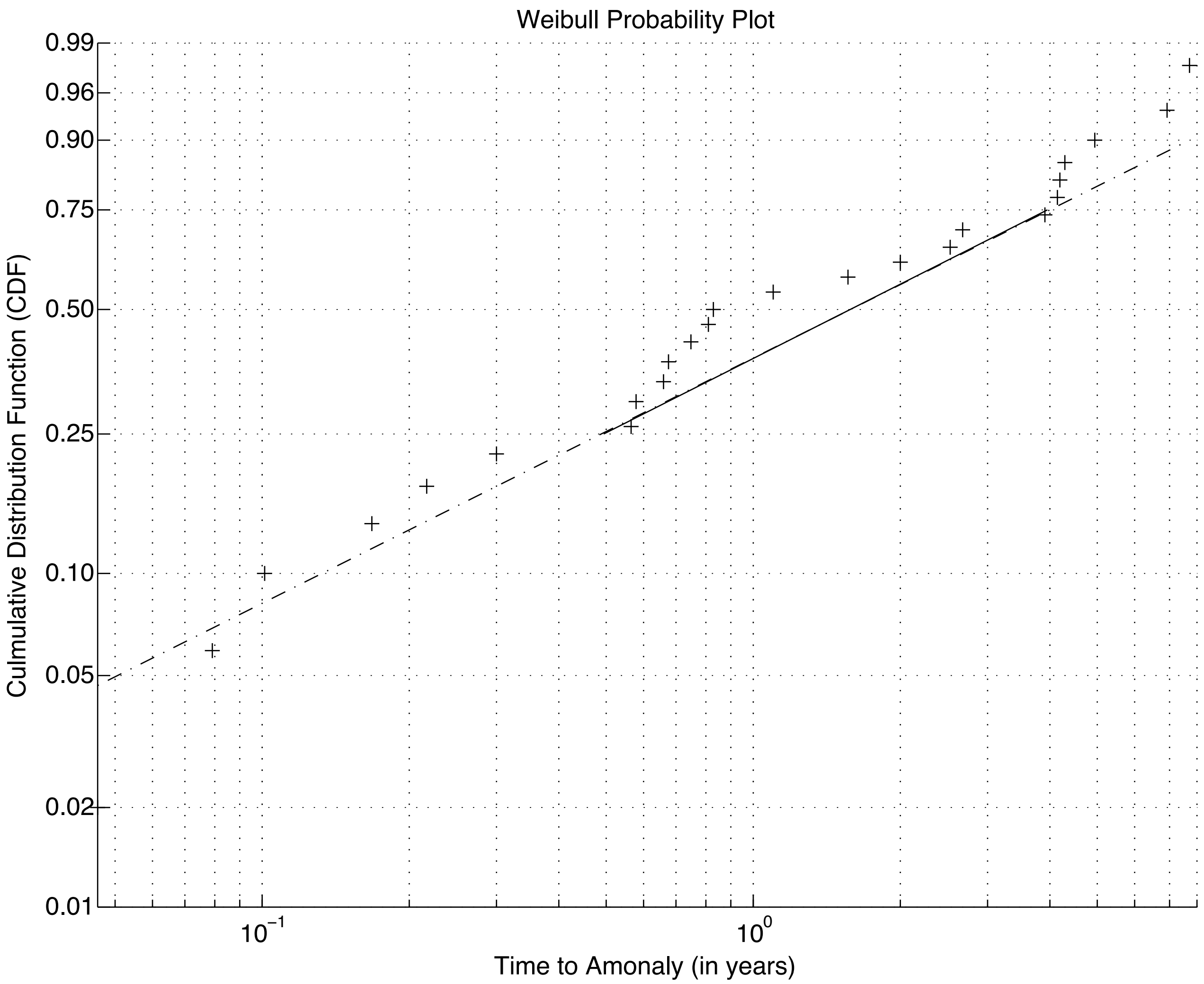

**Figure 3** (left panel): EP minor anomaly Weibull probability plot

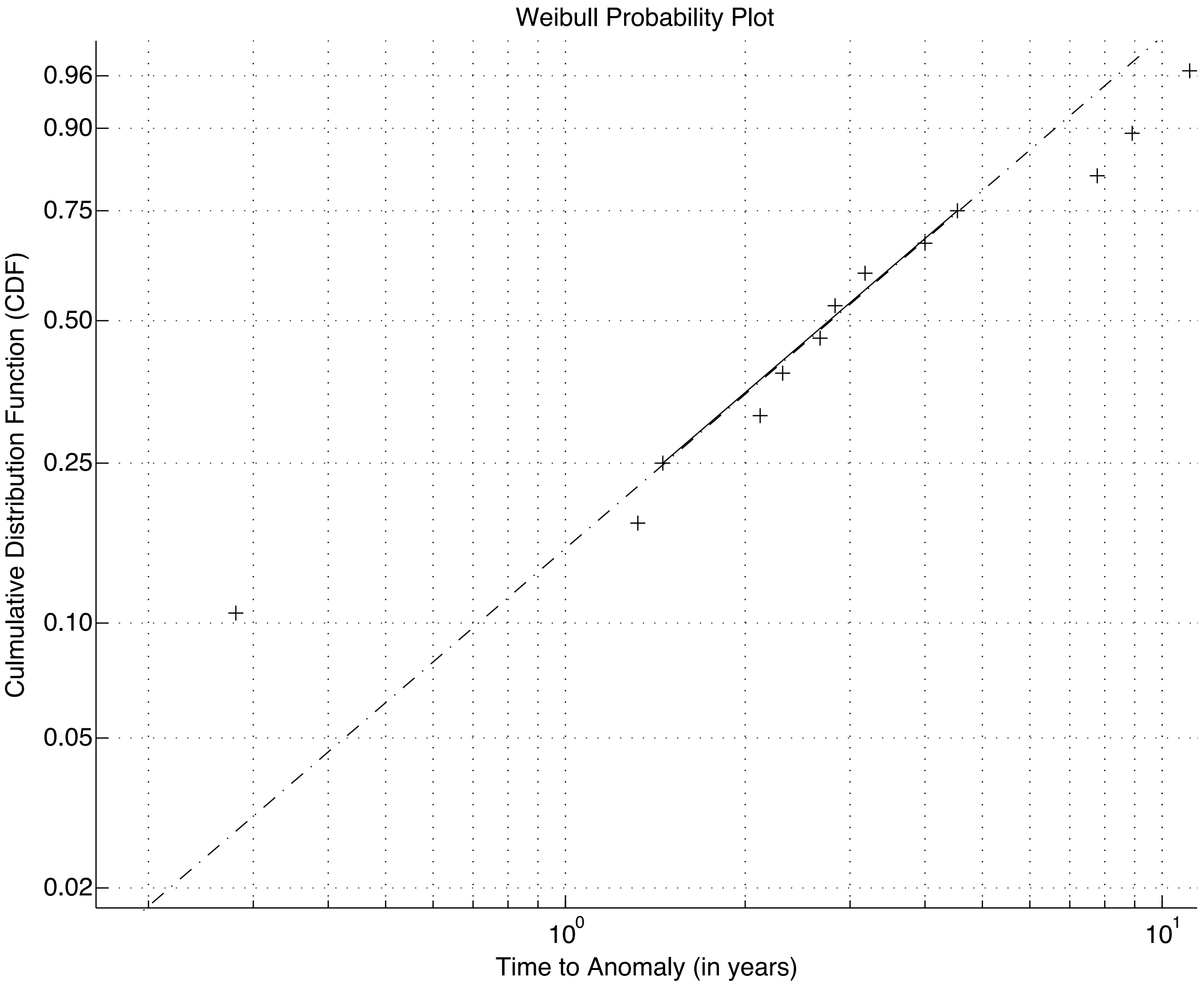

**Figure 3** (right panel): EP major anomaly Weibull probability plot